\documentclass[10pt,journal,compsoc]{IEEEtran}

\usepackage{cite}
\ifCLASSOPTIONcompsoc
    \usepackage[caption=false,font=footnotesize,labelfont=sf,textfont=sf]{subfig}
\else
    \usepackage[caption=false,font=footnotesize]{subfig}
\fi
\ifCLASSOPTIONcaptionsoff
    \usepackage[nomarkers]{endfloat}
\let\MYoriglatexcaption\caption
\renewcommand{\caption}[2][\relax]{\MYoriglatexcaption[#2]{#2}}
\fi
\usepackage[T1]{fontenc}
\usepackage{url}
\usepackage{subfig}
\usepackage{dsfont}
\usepackage{bm}
\usepackage[pdftex]{graphicx}
\usepackage{array,amsmath,amssymb,amsfonts}
\usepackage{amsthm}
\usepackage{subeqnarray}
\usepackage{cases}
\usepackage{booktabs}
\usepackage{color}
\usepackage[ruled,lined,linesnumbered,commentsnumbered,longend]{algorithm2e}
\makeatletter
\newcommand{\removelatexerror}{\let\@latex@error\@gobble}
\makeatother

\SetCommentSty{mycommfont}
\renewcommand{\vec}[1]{\boldsymbol{#1}}

\DeclareMathOperator*{\argmax}{argmax}
\DeclareMathOperator*{\diag}{diag}
\newtheorem{theorem}{\textbf{Theorem}}

\def\BibTeX{{\rm B\kern-.05em{\sc i\kern-.025em b}\kern-.08em
    T\kern-.1667em\lower.7ex\hbox{E}\kern-.125emX}}
\vspace{-0.3cm}
\IEEEoverridecommandlockouts

\usepackage{siunitx} % Required for alignment

\begin{document}

\title{Learning to Schedule Multi-Server Jobs with {\color{black}Fluctuated Processing Speeds}}

\author{
        Hailiang~Zhao,
        Shuiguang~Deng,~\IEEEmembership{Senior~Member,~IEEE,} 
        Feiyi~Chen,
        Jianwei~Yin,
        Schahram~Dustdar,~\IEEEmembership{Fellow,~IEEE},
        and~Albert~Y.~Zomaya,~\IEEEmembership{Fellow,~IEEE}%
\IEEEcompsocitemizethanks{
  \IEEEcompsocthanksitem H. Zhao, S. Deng, F. Chen, and J. Yin are with 
  the College of Computer Science and Technology, Zhejiang University, Hangzhou 310027, China.
  e-mail: \{hliangzhao, dengsg, chenfeiyi, zjuyjw\}@zju.edu.cn
  \IEEEcompsocthanksitem S. Dustdar is with the Distributed Systems Group, Technische Universität Wien, 1040 Vienna, Austria.
  e-mail: dustdar@dsg.tuwien.ac.at
  \IEEEcompsocthanksitem A. Y. Zomaya is with the School of Computer Science, University of Sydney, Sydney, 
  NSW 2006, Australia. e-mail: albert.zomaya@sydney.edu.au
  \IEEEcompsocthanksitem Shuiguang Deng is the corresponding author.}%
  % \thanks{Manuscript received November --, 2020; revised November --, 2020.}
}

\IEEEtitleabstractindextext{%
\begin{abstract}
Multi-server jobs are imperative in modern cloud computing systems. 
{\color{black}
A noteworthy feature of multi-server jobs is that, they usually request multiple computing devices simultaneously 
for their execution. How to schedule multi-server jobs online with a high system efficiency is a topic of great 
concern. Firstly, the scheduling decisions have to satisfy the service locality constraints. Secondly, the scheduling 
decisions needs to be made online without the knowledge of future job arrivals. Thirdly, and most importantly, the 
actual service rate experienced by a job is usually in fluctuation because of the dynamic voltage and frequency 
scaling (DVFS) and power oversubscription techniques when multiple types of jobs co-locate. A majority of online 
algorithms with theoretical performance guarantees are proposed. However, most of them require the processing speeds to 
be knowable, thereby the job completion times can be exactly calculated. To present a theoretically guaranteed online 
scheduling algorithm for multi-server jobs without knowing actual processing speeds apriori, in this paper, we propose 
\textsc{Esdp} (Efficient Sampling-based Dynamic Programming), which learns the distribution of the fluctuated processing speeds 
over time and simultaneously seeks to maximize the cumulative overall utility. The cumulative overall utility is 
formulated as the sum of the utilities of successfully serving each multi-server job minus the penalty on the operating, 
maintaining, and energy cost. \textsc{Esdp} is proved to have a polynomial complexity and a logarithmic regret, which is a 
State-of-the-Art result.} We also validate it with extensive simulations and the results show that the proposed 
algorithm outperforms several benchmark policies with improvements by up to 73\%, 36\%, and 28\%, respectively.
\end{abstract}

}

\maketitle

\IEEEdisplaynontitleabstractindextext

\IEEEpeerreviewmaketitle

\IEEEraisesectionheading{\section{Introduction}}\label{s1}
\IEEEPARstart{T}oday's computing clusters have plenty of multi-server jobs, e.g., the distributed training of deep 
neural networks \cite{zou2017distributed,gupta2018distributed} and large-scale graph computations \cite{yan2015effective,xiao2017tux2}. 
{\color{black}A notable feature of multi-server jobs is that they usually request multiple computing devices \textit{simultaneously} 
such as CPUs and GPUs and hold onto them during their execution.}
From Google cluster trace \cite{trace}, we can observe that more than 90\% jobs request multiple CPU cores and nearly 20\% jobs 
request CPU cores no less than 1000.

{\color{black}
It is difficult for the cluster scheduler to allocate an appropriate number of computing devices to each multi-server job 
with a high system efficiency. The major challenges are discussed as follows.
\begin{itemize}
    \item \textit{Service Locality.} Service locality is common in modern cloud and edge computing systems, especially for 
    \textit{Machine Learning as a Service} (MLaaS) \cite{ribeiro2015mlaas} and Serverless computing \cite{serverless1,serverless2}. 
    With service locality, a multi-server job may only be processed by a subset of servers where the computing device request, 
    software dependencies, and other requirements such as geographical constraints are satisfied. For instance, in a resource-constrained cluster, 
    service locality could lead to a situation where all the DNN training jobs are scheduled to the \textit{only} 
    server with GPUs and the rest of them have to wait until the GPUs are released.
    
    \item \textit{Unknown Arrival Patterns of Jobs.} In real-world scenarios, multi-server jobs arrive to the cluster online. 
    The scheduler needs to make the resource allocation decisions without knowing the job arrival patterns apriori. The 
    lack of the job arrival distributions could lead to the scheduling decision to \textit{a local optimum}.
    
    \item \textit{The Processing Speeds Experienced by each Job is Fluctuated.} In production systems where different 
    multi-server jobs co-locate, such as computation-intensive jobs, IO-intensive jobs, and 
    latency-critical jobs etc., the processing speeds may fluctuate over time and could be highly variable occasionally. 
    The reason is that the server is always in multi-tasking of different jobs, and the hardware techniques such as Dynamic Voltage and 
    Frequency Scaling (DVFS) \cite{wu2014green} and power oversubscription \cite{273871} adjust the CPU cycle frequency 
    constantly. 
\end{itemize}
}

A majority of online scheduling algorithms with theoretical guarantees have been proposed by formulating combinatorial 
optimization problems with scenario-oriented constraints 
\cite{gautam2015survey,BSP,8486422,attiya2020job,zhang2020evolving,liang2020data,narayanan2020heterogeneity,8941266,8486340,8737612,8737465,bao2018online,8917749}. 
To solve these combinatorial programs, algorithms are designed with various theoretical approaches. Typical approaches 
include relaxed integer programming \cite{BSP}, online primal-dual alternating updates \cite{8486422}, online approximate 
algorithms \cite{8737465,bao2018online,8917749}, heuristics \cite{attiya2020job,zhang2020evolving}, deep reinforcement 
learning (DRL) \cite{liang2020data,narayanan2020heterogeneity}, etc. {\color{black}However, despite the vast literature 
of them, their model formulations which tackle with the fluctuated processing speeds of multi-server jobs are limited. 
To execute these online algorithms, the processing speeds of servers are required to be \textit{knowable} when making the 
scheduling decisions, thereby the job completion times can be exactly calculated. However, as we have analyzed above, in 
production systems where different types of multi-server jobs co-locate, the actual processing speeds experienced by jobs 
is unknown and fluctuated when making the scheduling decisions.}

{\color{black}
To present a theoretically guaranteed online scheduling algorithm for multi-server jobs \textit{without knowing the distributions of the 
processing speeds apriori}, in this paper, we propose \textsc{Esdp} (Efficient Sampling-based Dynamic Programming) to learn the distributions of 
the fluctuated processing speeds with sufficient exploration-exploitation and simultaneously to maximize the cumulative overall utility (\textsc{Aou}). \textsc{Aou} is 
formulated as the sum of the obtained utilities of successfully processing each multi-server job minus the penalty on the operating, maintaining, and 
energy cost for serving them over each time slot. Further, the utility of a job is fitted by a \textit{stochastic quasi-linear} function of 
allocated computing devices in terms of its completion time. Our work is built on the intuition that, for a multi-server job, 
its completion time is mainly determined by the actual processing speed it experiences, which is linear with the allocated computing devices. 
Our basic assumption is that, although the actual processing speeds are fluctuated over time, \textit{they come from some certain distributions, which are determined 
by the hardware specifications of the underlying physical machines}. It is exactly \textsc{Esdp}'s job to \textit{learn} the underlying processing speed distributions 
and leverage them to guide the computing device allocations. Specifically, \textsc{Esdp} casts the online multi-server job scheduling problem into the framework of 
online learning \cite{online-learning-1}, and 
it makes the scheduling decisions for each arrived job with sufficient \textit{exploration-exploitation}. Based on the exploited patterns, 
\textsc{Esdp} introduces several deterministic maximization problems whose targets are the expectation of \textsc{Aou} approximated by statistics. Then, \textsc{Esdp} solves 
these deterministic problems with a dynamic programming subroutine in polynomial time. We use regret \cite{online-learning-1}, i.e., the gap on \textsc{Aou} 
between \textsc{Esdp}'s and \textit{the offline optimum} achieved by the oracle, to analyze the performance of \textsc{Esdp}. We provide a rigorous proof to show 
that \textsc{Esdp} has a best-so-far regret, i.e., $\mathcal{O}(\ln T)$, where $T$ is the time slot length. Our contribution fulfills one of the key deficiencies 
of current literature in the stochastic scheduling of mutli-server jobs without knowing processing speeds apriori. The main contributions are 
summarized as follows.
\begin{itemize}
    \item We propose an online algorithm, i.e., \textsc{Esdp}, to schedule multi-server jobs without exact processing speeds apriori. \textsc{Esdp} makes 
    no assumptions on the job arrival patterns, and it fully takes service locality into consideration. We prove that \textsc{Esdp} has a best-so-far 
    regret $\mathcal{O}(\ln T)$, which grows logarithmically with the time slot length.
    
    \item \textsc{Esdp} casts the online stochastic scheduling problem into the framework of online learning, and adopts several dynamic programming 
    subroutines to solve the approximated deterministic problems in polynomial time. 
    
    \item We validate the performance of \textsc{Esdp} with extensive simulations. Experimental results show that, in default settings, \textsc{Esdp} significantly 
    outperforms several widely used heuristics with improvements by up to 73\%, 36\%, and 28\%, respectively.
\end{itemize}
}

The rest of this paper is organized as follows. {\color{black}We formulate the stochastic multi-server job scheduling problem with a bipartite 
graph in Sec. \ref{s3}.} We then present the design details of \textsc{Esdp} with theoretical analysis in Sec. \ref{s4}. Numerical results are 
presented in Sec. \ref{s5}. We discuss related works in Sec. \ref{s2} and close this paper in Sec. \ref{s6}.
% ======================================================================================================================================================

\section{System Model}\label{s3}
We consider a computing cluster of heterogeneous servers serving several types of multi-server jobs. {\color{black}Different servers are equipped 
with different types and quantities of computing devices, including CPUs, GPUs, etc. Multi-server jobs of different types have different 
requests on computing devices under the service locality constraints.} Key notations used in this paper are summarized in Tab. \ref{notation}.

\begin{table}[htbp]   
    \color{black}
    \begin{center}
        \caption{\label{notation}Summary of key notations.}   
    \begin{tabular}{l|l}    
        \toprule
        {\textsc{Notation}}& {\textsc{Description}}\\[+0.1mm]
        \midrule
        $\mathcal{T}$ & Time horizon of length $T$\\[+0.7mm]
        $\mathcal{G} = (\mathcal{L}, \mathcal{R}, \mathcal{E})$ & The bipartite graph\\[+0.7mm]
        $l \in \mathcal{L}$ & A multi-server job type (port)\\[+0.7mm]
        $r \in \mathcal{R}$ & A server/node\\[+0.7mm]
        $(l,r) \in \mathcal{E}$ & The edge (channel) between $l$ and $r$\\[+0.7mm]
        $\forall r: \mathcal{L}_r$ & The set of job types connect to $r$\\[+0.7mm]
        $\forall l: \mathcal{R}_l$ & The set of servers connect to $l$\\[+0.7mm]
        $\rho_l(t)$ & The job arrival probability of port $l$ at time $t$\\[+0.7mm]
        $\mathds{1}_l(t) \in \{ 0, 1 \}$ & The job arrival status of port $l$ at time $t$\\[+0.7mm]
        $\mathcal{K}$ & The set of different types of computing devices\\[+0.7mm]
        $\vec{a}_k$ & The ovearll request on device $k$\\[+0.7mm]
        $c_k$ & The number of the type-$k$ devices in the cluster\\[+0.7mm]
        $\vec{x}(t)$ & The scheduling decision at time $t$\\[+0.7mm]
        $\forall k: c_k$ & The total number of type-$k$ devices\\[+0.7mm]
        $U_l(t)$ & The utility if job $l$ at time $t$\\[+0.7mm]
        $\forall k: f_k(\cdot)$ & Cost of provisioning type-$k$ devices\\[+0.7mm]
        $\textsc{U}\big(\vec{x}(t)\big)$ & The overall utility at time $t$\\[+0.7mm]
        $\textsc{Re}(T)$ & The regret over the time horizon $\mathcal{T}$\\[+0.7mm]
        \bottomrule   
    \end{tabular}  
    \end{center}
\end{table}

\subsection{Bipartite Graph Model under Service Localities}\label{s3.1}
We use a bipartite graph $(\mathcal{L}, \mathcal{R}, \mathcal{E})$ to model service locality, where $\mathcal{L}$ and $\mathcal{R}$ are 
the set of left vertices and right vertices, respectively, and $\mathcal{E}$ is the set of edges between the two sets of vertices. The 
vertices in $\mathcal{L}$ are indexed by $l$ and viewed as job types, while the vertices in $\mathcal{R}$ are indexed by $r$ and represent 
servers. For a vertex $l \in \mathcal{L}$, we use $\mathcal{R}_l \subseteq \mathcal{R}$ to represent the set of right vertices it connects 
with. Similarly, we use $\mathcal{L}_r \subseteq \mathcal{L}$ to represent the set of left vertices for $r \in \mathcal{R}$. 

\begin{figure}[htbp]
    \centerline{\includegraphics[width=3.3in]{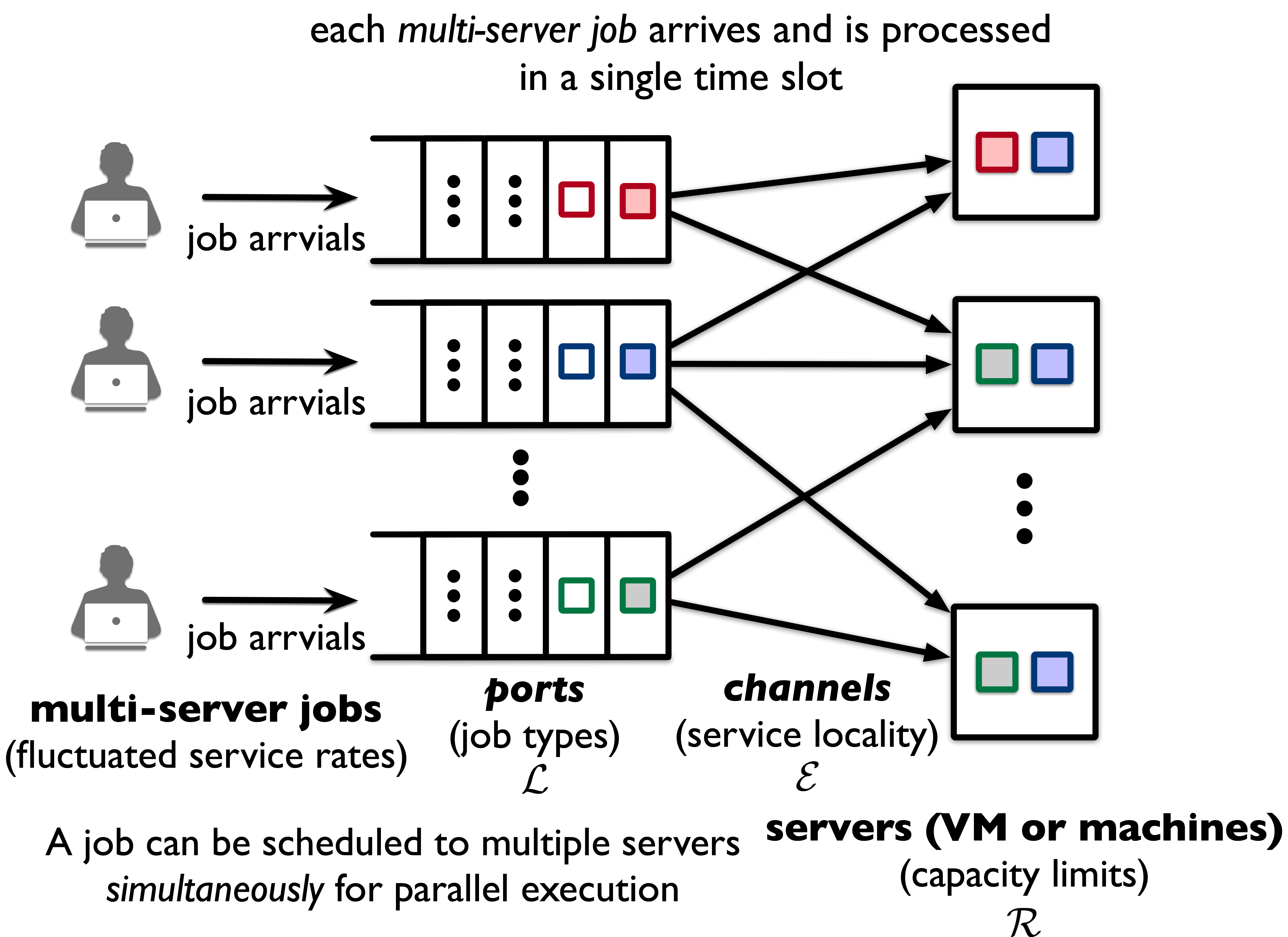}}
    \caption{\color{black}The bipartite graph model for multi-server job scheduling.}
    \label{fig1}
\end{figure}

We designate each vertex $l \in \mathcal{L}$ as \textit{port} and each edge $(l,r)$ as \textit{channel}. 
The bipartite graph model is visualized in Fig. \ref{fig1}.

\subsection{Job Scheduling with Restricted Capacities}\label{s3.2}
In our formulation, time is slotted, and at each time $t \in \mathcal{T} := \{1, ..., T\}$, for each port, at most one job 
arrives\footnote{\color{black}Even though, our model can be easily extended to the scenarios where multiple jobs 
arrive from a port in a single time slot.}. Concretely, at the beginning of time $t$, a job is yielded from port $l$ with probability 
$\rho_l(t)$, and with probability $1 - \rho_l(t)$, there is no job. 
{\color{black}It is worth noting that, the probabilities $\{ \rho_l(t) \}_{l \in \mathcal{L}}$ 
are only used for generating job arrival instances, which is not required by the to-be-proposed algorithm \textsc{Esdp} when making the online decisions.}

There are $K$ types of computing devices in the cluster, including CPUs, GPUs, NPUs, and FPGAs. For each type-$l$ job, we denote by 
$a_k^{(l,r)} \in \mathbb{N}^+$ its request on the type-$k$ computing device when it is processed by server $r$ through the channel $(l,r)$. 
The total number of the type-$k$ computing devices in the cluster, where $k \in \mathcal{K} := \{1, .., K\}$, is represented by 
$c_k \in \mathbb{N}^+$.  

At time $t$, we use 
\begin{equation}
    \vec{x}(t) := \Big[x_{(l,r)}(t)\Big]^\mathrm{T}_{\forall (l,r) \in \mathcal{E}} \in \mathcal{X} := \big\{0, 1\big\}^{|\mathcal{E}|}
\end{equation}
to represent the scheduling decision. A job can be scheduled to multiple servers simultaneously for parallel execution. 
A constraint $\vec{x}(t)$ should satisfy is that, the computing devices allocated out from the cluster should not more than it has:
\begin{equation}
    \sum_{(l, r) \in \mathcal{E}} a_k^{(l, r)} x_{(l,r)}(t) \leq c_k, \forall k \in \mathcal{K}, t \in \mathcal{T}.
    \label{cons1}
\end{equation}
Note that if port $l$ yields no job at $t$, denoted by $\mathds{1}_l(t) = 0$, then $x_{(l,r)}(t) = 0$ for all $r \in \mathcal{R}_l$. 

{\color{black}
The multi-server job scheduling problem is studied for maximizing the \textsc{Aou}, which is formulated as the sum 
of the utilities of successfully serving each multi-server job minus the penalty on the operating, maintaining, and energy cost for 
serving them over each time slot. We denote by $U_l(t)$ the utility of the type-$l$ job at time $t$, and it is formulated as
\begin{flalign}
    U_l(t) := \sum_{r \in \mathcal{R}_l} x_{(l,r)}(t) Z_{(l,r)} (t) - 
    \sum_{k \in \mathcal{K}} \sum_{r \in \mathcal{R}_l} f_k \Big(a_k^{(l,r)} \Big) x_{(l,r)} (t),
    \label{Ut}
\end{flalign}
where $Z_{(l,r)} (t)$ is a stochastic variable following an underlying distribution with the expectation of $\upsilon_{(l,r)}$. 
We formulate $Z_{(l,r)} (t)$ as the actual computation utility experienced by the type-$l$ job at time $t$ when it is 
processed by server $r$ through the channel $(l,r) \in \mathcal{E}$. Correspondingly, $\upsilon_{(l,r)}$ is the expectation 
of the computation utility, and it is unknown when making the scheduling decisions. Our formulation is built on the assumption 
that, although $\upsilon_{(l,r)}$ cannot be obtained apriori, we can learn it and approximate it with sufficient 
\textit{exploration-exploitation}. In addition, the computation utility is \textit{linearly additive}, i.e., 
if a job is processed through multiple channels in parallel, the final utility is the sum of computation utility obtained from all 
these channels.}
The second part in \eqref{Ut} is the penalty on the supply cost. Thereinto, $f_k \big(a_k^{(l,r)} \big)$ is the supply cost for 
provisioning $a_k^{(l,r)}$ units of the type-$k$ computing device for the type-$l$ job through the channel $(l,r)$. 
$\{f_k (\cdot) \}_{\forall k \in \mathcal{K}}$ models the operating, maintaining, and energy cost for serving 
jobs. Different from previous works \cite{dpos-base,dpos,onsocmax}, we make no assumptions on the convexity or differentiability of 
$\{f_k (\cdot) \}_{\forall k \in \mathcal{K}}$.

Our goal is to maximize the expectation of \textsc{Aou}, i.e., the expected sum of utilities of multi-server jobs in a long-term horizon. 
The problem is formulated as follows. 
{
\color{black}
\begin{flalign}
    \mathcal{P}_1: &\qquad \max_{\forall t \in \mathcal{T}: 
    \vec{x}(t) \in \mathcal{X}} \lim_{T \to \infty} \sum_{t = 1}^T \mathbb{E}\bigg[ \sum_{l \in \mathcal{L}} U_l(t) \bigg] \nonumber \\
    s.t. &\qquad \qquad \qquad \qquad \quad \eqref{cons1}, \nonumber\\ 
    &\sum_{r \in \mathcal{R}_l} x_{(l,r)}(t) = 0 \text{ \textit{if} } \mathds{1}_l(t) = 0, \forall l \in \mathcal{L}, t \in \mathcal{T}, \label{cons2}
\end{flalign}
}
With further transformation, we can get
\begin{equation}
    \mathbb{E}\bigg[ \sum_{l \in \mathcal{L}} U_l(t) \bigg] = \sum_{(l,r) \in \mathcal{E}} x_{(l,r)} (t) 
    \bigg[ Z_{(l,r)}(t) - \sum_{k \in \mathcal{K}} f_k\big(a_k^{(l,r)}\big) \bigg].
\end{equation}
In the following content, we use $\textsc{U}\big(\vec{x}(t)\big)$ and $\sum_{l \in \mathcal{L}} U_l(t)$ interchangeably. 
% ======================================================================================================================================================

\section{Algorithm Design}\label{s4}
{\color{black}
In this section, we demonstrate the design details of \textsc{Esdp}, which can solve $\mathcal{P}_1$ with a \textit{probabilistic} optimality 
asymptotically in polynomial time. \textsc{Esdp} is built on the well known ESCB policy \cite{ESCB,ESCB2} and a recent derivative, called 
AESCB \cite{semi-bandit}, for solving combinatorial semi-bandit problems. In the following content, firstly, we formulate the regret 
minimization problem that corresponds to $\mathcal{P}_1$ and bring in several evolving statistics to approximate 
$\mathbb{E}\big[ \textsc{U} \big(\vec{x}(t) \big) \big]$ at each time $t$. Based on these statistics and a converge-to-zero sequence 
$\{\delta(t)\}_{t \in \mathcal{T}}$, we introduce a series of deterministic optimization problems. Then, we solve these deterministic 
problems sequentially based on dynamic programming in polynomial times. After that, we provide rigorous 
theoretical analysis for \textsc{Esdp} in terms of the algorithmic complexity and the regret on \textsc{Aou}. In the end, we discuss the possible 
extensions of \textsc{Esdp} on the Gang scheduling scenarios. 
}

\subsection{Regret Minimizing with Evolving Statistics}\label{s4.1}
$\mathcal{P}_1$ is an online \textit{stochastic} optimization problem with random variables 
$\vec{Z}(t) = [Z_{(l,r)} (t)]^\textrm{T}_{\forall (l,r) \in \mathcal{E}}$ not determined until the time $t$ arrives. $\mathcal{P}_1$ is 
equivalent to the regret minimization problem listed below:
\begin{flalign}
    \mathcal{P}_2: \min_{\forall t \in \mathcal{T}: \vec{x}(t) \in \mathcal{X}} \lim_{T \to \infty} \textsc{Re}(T) := 
    \sum_{t=1}^T \mathbb{E}\Big[ \Delta\big( \vec{x}(t) \big) \Big] \quad \nonumber \\
    s.t. \quad \eqref{cons1}, \eqref{cons2}, \qquad \qquad \qquad \qquad \quad \nonumber
\end{flalign}
where the expected per-time slot gap $\mathbb{E}\big[ \Delta \big( \vec{x}(t) \big) \big]$ is 
\begin{equation}
    \mathbb{E}\Big[ \Delta\big( \vec{x}(t) \big) \Big] := \tilde{\vec{\upsilon}}^\mathrm{T} \vec{x}^*(t) - 
    \mathbb{E}\Big[ \sum_{l \in \mathcal{L}} U_l(t) \Big]
    \label{gap}
\end{equation}
and 
\begin{equation}
    \left\{
        \begin{array}{l}
            \tilde{\vec{\upsilon}} := \big[\upsilon_{(l,r)} - 
            \sum_{k \in \mathcal{K}} f_k\big(a_k^{(l,r)}\big)\big]^\mathrm{T}_{\forall (l,r) \in \mathcal{E}} \in [0,1]^{|\mathcal{E}|} \\
            
            \vec{x}^*(t) := \argmax_{\vec{x}(t) \in \Omega(t)} \big\{ \tilde{\vec{\upsilon}}^\mathrm{T} \vec{x}(t) \big\} \\
            
            \Omega(t) := \big\{\vec{x}(t) \in \mathcal{X} \mid \eqref{cons1} \text{ \textrm{\&} } \eqref{cons2} \text{ hold at time }t \big\}.
        \end{array}
    \right.
\end{equation}
{\color{black}
The regret $\textsc{Re}(T)$ is the gap between the optimal \textsc{Aou} achieved by an \textit{omniscient} oracle who has the full 
knowledge on $\vec{\upsilon}$ and the \textsc{Aou} achieved by the to-be-proposed algorithm \textsc{Esdp}. A good algorithm should achieve a 
smallest possible regret $\textsc{Re}(T)$ as $T$ goes to infinity.}
For simplification, we denote by $\tilde{\vec{Z}}(t)$ the column vector 
\begin{equation*}
    \bigg[Z_{(l,r)}(t) - \sum_{k \in \mathcal{K}} f_k(a_k^{(l,r)})\bigg]^\mathrm{T}_{\forall (l,r) \in \mathcal{E}}.
\end{equation*}
{\color{black}W.O.L.G, we normalize $\tilde{\vec{Z}}(t)$ into $[0, 1]^{|\mathcal{E}|}$ by carefully tuning the parameters in $\{f_k (\cdot)\}_{k \in \mathcal{K}}$.}
The non-negative property is widely accepted for utility functions 
\cite{utility,dpos-base,onsocmax,assume3}. Nevertheless, different from the above literature, we make no assumptions on the convexity or 
differentiability of $\{f_k\}_{\forall k \in \mathcal{K}}$.

{\color{black}$\mathcal{P}_2$ is still a stochastic optimization problem and the expectation operation is not eliminated. To make it solvable, based on the idea 
introduced by the ESCB policy \cite{ESCB}, we introduce several statistics to approximate $\vec{\upsilon}$ with the explorated information.}
These statistics are used to supersede the random variables in $\mathcal{P}_2$. Specifically, at each time $t$, we define 
\begin{equation}
    n_{(l,r)}(t) := \sum_{t'=1}^t x_{(l,r)} \big(t'\big)
\end{equation}
as the \textit{cumulative quantity} of channel $(l,r) \in \mathcal{E}$ been used up to time $t$. Based on it, we define the following statistics:
\begin{flalign}
    \hat{\upsilon}_{(l,r)}(t) &:= \left\{
        \begin{array}{ll}
            \frac{\sum_{t'=1}^t x_{(l,r)}(t') \tilde{Z}_{(l,r)}(t')}{n_{(l,r)}(t)} & n_{(l,r)}(t) > 0 \\
            0 & \textrm{otherwise}
        \end{array}
    \right. \label{st1} \\
    \hat{\sigma}_{(l,r)}^2(t) &:= \left\{
        \begin{array}{ll}
            \frac{g(t)}{2 n_{(l,r)}(t)} & n_{(l,r)}(t) > 0 \\
            +\infty & \textrm{otherwise},
        \end{array}
    \right. \label{st2}
\end{flalign}
where
\begin{equation}
    g(t) := \ln t + 4 \ln (\ln t + 1) \cdot \max_{t' \in \mathcal{T}} \bigg\{\max_{\vec{x} \in \Omega(t')} \| \vec{x} \|_1 \bigg\}.
\end{equation}
$\hat{\upsilon}_{(l,r)}(t)$ is a non-biased estimation based on historical noisy computation utilities for type-$l$ job when processed through channel $(l,r)$. 
$\hat{\sigma}_{(l,r)}^2(t)$ is a metric proportional to the variance of the estimate $\hat{\upsilon}_{(l,r)}(t)$, proposed by \cite{ESCB}. 
We place a hat on the estimations to indicate that they are calculated and updated online. 

Inspired by the ESCB and AESCB policies, at time $t$, we introduce the following \textit{deterministic} problem $\mathcal{P}_3(t)$:
\begin{flalign}
    \mathcal{P}_3(t): \max_{\vec{x}(t) \in \Omega(t)} &\tilde{\textsc{U}}(\vec{x}(t)) := 
    \delta(t) + \hat{\vec{\upsilon}}(t)^\textrm{T} \vec{x}(t) + \sqrt{\hat{\vec{\sigma}}^2(t)^\textrm{T} \vec{x}(t)} \nonumber \\
    s.t. &\qquad \qquad \quad \eqref{cons1}, \nonumber\\ 
    &\quad \delta(t) > 0, \lim_{t \to \infty} \delta(t) = 0, \label{delta(t)}
\end{flalign}
where 
\begin{equation*}
    \left\{
        \begin{array}{l}
            \hat{\vec{\upsilon}}(t) := \Big[\hat{\upsilon}_{(l,r)}(t)\Big]_{(l,r) \in \mathcal{E}}^\mathrm{T} \\
            \hat{\vec{\sigma}}^2(t) := \Big[\hat{\sigma}_{(l,r)}^2(t)\Big]_{(l,r) \in \mathcal{E}}^\mathrm{T}
        \end{array}
    \right.
\end{equation*}
are the corresponding column vectors. 
{\color{black}
Moreover, $\hat{\vec{\upsilon}}(t)$ can be efficiently calculated through matrix operations as follows:
\begin{equation*}
    \phi \bigg( \big[ \vec{x}(1), ..., \vec{x}(t) \big] 
    \Big[ \big(\tilde{\mathbf{Z}}(1) \oslash \vec{n}(1) \big)^\mathrm{T}, ..., \big(\tilde{\mathbf{Z}}(t) \oslash \vec{n}(t) \big)^\mathrm{T} \Big]^\mathrm{T} \bigg),
\end{equation*}
where $\oslash$ is the element-wise division operator, $\vec{n}(t)$ is the vector $\{n_{(l,r)}(t)\}_{(l,r)\in \mathcal{E}}^\mathrm{T}$, and $\phi(\cdot)$ is the inverse of function $\diag(\cdot)$, defined as
\begin{equation}
    \phi ( \mathbf{M} ) := \sum_{i = 1}^{|\mathcal{E}|} \big( \vec{e}_i^\mathrm{T} \mathbf{M} \vec{e}_i \big) \vec{e}_i, \quad  \mathbf{M} \in \mathbb{R}^{|\mathcal{E}| \times |\mathcal{E}|}.
    \label{phi}
\end{equation} 
In \eqref{phi}, $\vec{e}_i$ is the $i$-th standard unit basis. 
}

In $\mathcal{P}_3(t)$, $\{\delta(t)\}_{t \in \mathcal{T}}$ could be any sequence converges to zero. 
For instance,
\begin{equation}
    \delta(t) := \frac{1}{\ln \Big(\ln t + 1 \Big) + 1}.
\end{equation}
{\color{black}The objective of $\mathcal{P}_3(t)$ is an approximated \textit{statistical-based} overall computation utility at time $t$. 
From $\mathcal{P}_2$ to $\mathcal{P}_3(t)$, we remove the random variable 
$\vec{Z}(t)$ and thereby remove the expectation operation in the objective. As a result, we transform 
the original stochastic problem into a deterministic problem while keeping the solution space impervious.}
In most case, the following inequality should hold:
\begin{equation}
    \bigg| \Big(\tilde{\vec{\upsilon}} - \hat{\vec{\upsilon}}(t) \Big)^\mathrm{T} \vec{x}(t) \bigg| \leq \sqrt{\hat{\vec{\sigma}}^2(t)^\mathrm{T}\vec{x}(t)}.
\end{equation}
By Chebyshev's Inequality, $\hat{\vec{\upsilon}}(t)^\mathrm{T} \vec{x}(t) \pm  \sqrt{\hat{\vec{\sigma}}^2(t)^\mathrm{T}\vec{x}(t)}$ covers 
nearly $60\%$ population. To achieve a larger coverage, we can increase the numerical multiplier to the standard variance. In our formulation, 
setting the multiplier as $1$ is enough to achieve the State-of-the-Art minimum regret upper bound. The analysis will be detailed in Sec. \ref{s4.3}.

\subsection{Polynomial-time Dynamic Programming}\label{s4.2}
If the sequence $\{\delta(t)\}_{t \in \mathcal{T}}$ is removed from $\tilde{\textsc{U}}(\vec{x}(t))$ and \eqref{delta(t)} is dropped, 
$\mathcal{P}_3(t)$ is NP-hard \cite{ESCB,ESCB2}, i.e., it cannot be solved in polynomial time. 
{\color{black}Therefore, to solve it efficiently, inspired by the 
AESCB policy \cite{semi-bandit}, \textsc{Esdp} resorts to solving several \textit{relaxed} budgeted integer programming problems by adding the 
converge-to-zero sequence $\{\delta(t)\}_{t \in \mathcal{T}}$, which is exactly what we have done when formulating $\mathcal{P}_3(t)$.}

In the following, we will detail how we solve $\mathcal{P}_3(t)$ with dynamic programming. 
Firstly, at each time $t$, based on $\delta(t)$, we define the following scale-up 
statistics for $\hat{\upsilon}_{(l,r)}(t)$ and $\hat{\sigma}_{(l,r)}^2(t)$ respectively:
\begin{flalign}
    \hat{\Upsilon}_{(l,r)} (t) &:= \Big\lceil \xi(t) \hat{\upsilon}_{(l,r)}(t) \Big\rceil \label{scal-mean}\\
    \hat{\Sigma}_{(l,r)}^2 (t) &:= \Big\lceil \xi^2(t) \hat{\sigma}_{(l,r)}^2(t) \Big\rceil, \label{scal-var}
\end{flalign}
where 
\begin{equation}
    \xi(t) := \Bigg\lceil \frac{ \max_{t' \in \mathcal{T}} \big\{\max_{\vec{x} \in \Omega(t')} \| \vec{x} \|_1 \big\} }{\delta(t)} \Bigg\rceil
    \label{delta-defi}
\end{equation}
is the scaling size at time $t$. By the AESCB policy \cite{semi-bandit}, at each time $t$, we introduce several budgeted integer programming problems 
$\mathcal{P}_4(s,t)$ for each $s \in \mathcal{S}(t)$, where 
\begin{equation}
    \mathcal{S}(t) := \bigg\{0, 1, ..., \xi(t) \cdot \max_{t' \in \mathcal{T}} \max_{\vec{x} \in \Omega(t')} \| \vec{x} \|_1 \bigg\},
\end{equation}
as follows: 
\begin{flalign}
    \mathcal{P}_4(s, t): \qquad \max_{\vec{x}(t) \in \mathcal{X}} \hat{\vec{\Sigma}}^2 (t)^\mathrm{T} \vec{x}(t) \qquad \qquad \quad \nonumber \\
    s.t. \qquad \quad \eqref{cons1}, \eqref{delta(t)}, \qquad \qquad \qquad \quad \nonumber \\
    \hat{\vec{\Upsilon}} (t)^\mathrm{T} \vec{x}(t) \geq s. \qquad \qquad \quad \label{s}
\end{flalign}
In $\mathcal{P}_4(s, t)$, $\hat{\vec{\Sigma}}^2 (t)$ and $\hat{\vec{\Upsilon}} (t)$ are the corresponding column vectors for \eqref{scal-mean} and 
\eqref{scal-var}, respectively. Let us use $\vec{x}_{\mathcal{P}_4}^*(s,t)$ to denote the optimal solution for $\mathcal{P}_4(s,t)$. Then, the 
final solution to $\max\{\mathcal{P}_4(s,t)\}_{s \in \mathcal{S}(t)}$ at time $t$, denoted by $\vec{x}_{\mathcal{P}_4}^*(t)$, is set as some 
$\vec{x}_{\mathcal{P}_4}^*(s^\star,t)$ where $s^\star \in \mathcal{S}(t)$ staisfies 
\begin{equation}
    s^\star \in \argmax_{s \in \mathcal{S}(t)} \Bigg\{ s + \sqrt{\hat{\vec{\Sigma}}^2 (t)^\mathrm{T} \vec{x}_{\mathcal{P}_4}^*(s,t)} \Bigg\}.
    \label{P4}
\end{equation}
{\color{black}Now we demonstrate the detailed procedure of \textsc{Esdp}, which is summarized in \textbf{Algorithm \ref{algo1}}. \textsc{Esdp} solves $\mathcal{P}_1$ and 
$\mathcal{P}_2$ by solving the problems $\{ \mathcal{P}_4(s, t) \}_{s \in \mathcal{S}(t), t \in \mathcal{T}}$. 
The relations between $\mathcal{P}_3(t)$ and $\{\mathcal{P}_4(s,t)\}_{s \in \mathcal{S}(t)}$, and how the solutions of 
$\{\mathcal{P}_4(s,t)\}_{s \in \mathcal{S}(t), t \in \mathcal{T}}$ affect the regret $\textsc{Re}(T)$ will be analyzed in Sec. \ref{s4.3}.}

\begin{figure}
    \removelatexerror
    \begin{algorithm}[H]
        \label{algo1}
        \caption{The \textsc{Esdp} Framework}
        \KwIn{The bipartite graph $(\mathcal{L}, \mathcal{R}, \mathcal{E})$, requirements $\{a_k^{(l,r)}\}_{k \in \mathcal{K}, (l,r) \in \mathcal{E}}$, 
        capacities $\{c_k\}_{k \in \mathcal{K}}$, cost functions $\{f_k\}_{k \in \mathcal{K}}$, and the sequence $\{\delta(t)\}_{t\in \mathcal{T}}$}
        \KwOut{Online solution to $\mathcal{P}_1$ (and $\mathcal{P}_2$) at time $t \in \mathcal{T}$}
        \While{$t = 1, ..., T$}
        {
            Observe the job arrival status from each port $l \in \mathcal{L}$\\
            Update $\hat{\vec{\Upsilon}}(t)$ and $\hat{\vec{\Sigma}}^2 (t)$ with \eqref{scal-mean} and \eqref{scal-var} based on $\delta(t)$, respectively \\
            \tcc{Solve $\{\mathcal{P}_4(s,t)\}_{s \in \mathcal{S}(t)}$ by Algorithm \ref{algo2}}
            \For{each $s \in \mathcal{S}(t)$}
            {
                Solve $\mathcal{P}_4(s,t)$ and return $\vec{x}_{\mathcal{P}_4}^*(s,t)$\\
            }
            $\vec{x}_{\mathcal{P}_4}^*(t) \leftarrow \vec{x}_{\mathcal{P}_4}^*(s^\star, t)$, where $s^\star$ staisfies \eqref{P4}\\
            \tcc{Satisfy constraint \eqref{cons2} of $\mathcal{P}_1$}
            \For{each $l \in \mathcal{L}$}
            {
                \If{$\mathds{1}_l(t) == 0$}
                {
                    \For{each $r \in \mathcal{R}_l$}
                    {
                        Set the $(l,r)$-th element of $\vec{x}_{\mathcal{P}_4}^*(t)$ as $0$\\
                    }
                }
            }
        }
        \Return{$\big\{\vec{x}_{\mathcal{P}_4}^*(t)\big\}_{t\in\mathcal{T}}$ and $\big\{\textsc{U}\big(\vec{x}_{\mathcal{P}_4}^*(t)\big)\big\}_{t \in \mathcal{T}}$}
    \end{algorithm}
\end{figure}

Now, the problem is how to solve $\{\mathcal{P}_4(s,t)\}_{s \in \mathcal{S}(t)}$ optimally within polynomial time. \textsc{Esdp} solves it based on 
dynamic programming. Concretely, at each time $t$, corresponding to each $\mathcal{P}_4(s,t)$, we bring in the problem $\mathcal{P}_5(s,t,\vec{c},i)$ 
as follows.
\begin{flalign}
    \mathcal{P}_5(s,t,\vec{c},i): \quad \max_{\vec{x}(t) \in \mathcal{X}} \hat{\vec{\Sigma}}^2 (t)^\mathrm{T} \vec{x}(t) \qquad \qquad \quad \nonumber \\
    s.t. \qquad \eqref{cons1}, \eqref{delta(t)}, \eqref{s}, \qquad \qquad \qquad \nonumber \\
    \sum_{e = e_1}^{e_i} x_e(t) = 0, \qquad \qquad \qquad \label{force}
\end{flalign}
where $\vec{c} := [c_k]^\mathrm{T}_{k \in \mathcal{K}}$ is the capacity vector in \eqref{cons1}, $e := (l,r) \in \mathcal{E}$ and 
$e_i$ is the $i$-th edge $(l,r)$ in $\mathcal{E}$. The new constraint \eqref{force} is used to set the first several scheduling decisions (until $i$) 
to $0$ forcibly. Obviously, $\mathcal{P}_5(s,t,\vec{c},0)$ is equal to $\mathcal{P}_4(s,t)$ because \eqref{force} is not functioning when $i=0$. 
The optimal solution of $\mathcal{P}_5(s,t,\vec{c},i)$ can be obtained by recursing over $s$, $\vec{c}$, and $i$. To do this, let us use 
$\vec{x}^*(s,t,\vec{c},i)$ to denote the optimal solution of $\mathcal{P}_5(s,t,\vec{c},i)$, and use $V_{\mathcal{P}_5}^*(s,t,\vec{c},i)$ to 
denote the corresponding objective. In the following, we demonstrate the recursing details.

\underline{\textbf{Case I}}: If $x_{e_{i+1}}^*(s,t,\vec{c},i) = 0$, i.e., the $(i+1)$-element of $\vec{x}^*(s,t,\vec{c},i)$ is $0$, then \eqref{force} 
is not violated for $\mathcal{P}_5(s, t, \vec{c}, i+1)$. Thus, we have 
\begin{flalign}
    \vec{x}^*(s,t,\vec{c},i+1) = \vec{x}^*(s,t,\vec{c},i)
\end{flalign}
and 
\begin{flalign}
    V_{\mathcal{P}_5}^*(s,t,\vec{c},i+1) = V_{\mathcal{P}_5}^*(s,t,\vec{c},i).
\end{flalign}
The result means that $\vec{x}^*(s,t,\vec{c},i)$ is also the optimal solution to $\mathcal{P}_5(s, t, \vec{c}, i+1)$.

\underline{\textbf{Case II}}: If $x_{e_{i+1}}^*(s,t,\vec{c},i) = 1$, the optimal substructure is much more complicated. For simplification, we define 
matrix $\mathbf{A}$ by 
\begin{equation*}
    \mathbf{A} = \Big[a_k^{(l,r)}\Big]^{K \times |\mathcal{E}|}.
\end{equation*}
Then we have
\begin{equation}
    \mathbf{A}\Big(\vec{x}^*(s,t,\vec{c},i) - \vec{e}_{i+1}\Big)\leq \vec{c} - A_{:, i+1},
    \label{p1}
\end{equation}
where $\vec{e}_{i+1}$ is the $(i+1)$-th standard unit basis. Besides, 
\begin{flalign}
    \hat{\vec{\Upsilon}}(t)^\mathrm{T} \Big(\vec{x}^*(s,t,\vec{c},i) - \vec{e}_{i+1} \Big) \geq s - \hat{\Upsilon}_{e_{i+1}}(t)
    \label{p2}
\end{flalign}
and
\begin{equation*}
    \hat{\vec{\Sigma}}^2 (t)^\mathrm{T} \big(\vec{x}^*(s,t,\vec{c},i) - \vec{e}_{i+1}\big) = 
    \hat{\vec{\Sigma}}^2 (t)^\mathrm{T} \vec{x}^*(s,t,\vec{c},i) - \hat{\Sigma}^2_{e_{i+1}}(t).
\end{equation*}
Combining the above formula with \eqref{p1} and \eqref{p2}, we can get the following evolving optimal substructure:
\begin{flalign}
    V_{\mathcal{P}_5}^*(s,t,\vec{c},i) = &V_{\mathcal{P}_5}^*\Big(\max \Big\{s - \hat{\Upsilon}_{e_{i+1}}(t), 0 \Big\}, t, \nonumber \\
    &\max \{\vec{c} - A_{:, i+1}, 0\} ,i+1\Big) + \hat{\Sigma}^2_{e_{i+1}}(t).
\end{flalign}
Thus, for every possible $s$, $\vec{c}$, and $i$, we can update the solution to $\mathcal{P}_5(s,t,\vec{c},i)$ by 
\begin{equation*}
    x_{e_{i+1}}^*(s,t,\vec{c},i) = \left\{
    \begin{array}{ll}
        0 & V_{\mathcal{P}_5}^*(s,t,\vec{c},i) = V_{\mathcal{P}_5}^*(s,t,\vec{c},i+1)\\
        1 & \text{otherwise}.
    \end{array}
    \right.
\end{equation*}

\begin{figure}
    \removelatexerror
    \begin{algorithm}[H]
        \label{algo2}
        \caption{DP for solving $\{\mathcal{P}_4(s,t)\}_{s\in\mathcal{S}(t)}$}
        \KwIn{$\mathcal{S}(t)$, resource requirements $\{a_k^{(l,r)}\}_{k \in \mathcal{K}, (l,r) \in \mathcal{E}}$, and scale-up statistics 
        $\hat{\vec{\Upsilon}}(t)$ and $\hat{\vec{\Sigma}}(t)$}
        \KwOut{Optimal solution to $\{\mathcal{P}_4(s,t)\}_{s\in\mathcal{S}(t)}$}
        $\forall i \in \{ 0, ..., |\mathcal{E}| \}$, $\vec{x}^*(s,t,\vec{c},i) \leftarrow \vec{0}$
        \For{$s$ from $0$ to $\xi(t) \cdot \max_{t' \in \mathcal{T}} \big\{\max_{\vec{x} \in \Omega(t')} \| \vec{x} \|_1 \big\}$}
        {
            \For{$\vec{c'}$ from $\vec{0}$ to $\vec{c}$}
            {
                $V_{\mathcal{P}_5}^*(s,t,\vec{c}',|\mathcal{E}|)$ \textbf{is} $0$ \textbf{if} $s == 0$ \textbf{else} $-\infty$ \\
                \For{$i$ from $|\mathcal{E}|-1$ to $0$}
                {
                    \If{$\vec{c}' == \vec{0}$}
                    {
                        $V_{\mathcal{P}_5}^*(s,t,\vec{c}',i) \leftarrow V_{\mathcal{P}_5}^*(s,t,\vec{c}',i+1)$ \\
                        \textbf{continue}\\
                    }
                    $V_{\mathcal{P}_5}^*(s,t,\vec{c}',i) \leftarrow 
                    \max \bigg\{ V_{\mathcal{P}_5}^*\Big(\max \big\{s - \hat{\Upsilon}_{e_{i+1}}(t), 0 \big\}, t, \max \{ \vec{c}' - A_{:, i+1}, 0 \},i+1 \Big) + \hat{\Sigma}^2_{e_{i+1}}(t), 
                    V_{\mathcal{P}_5}^*(s,t,\vec{c}',i+1) \bigg\}$ \\
                    \If{$V_{\mathcal{P}_5}^*(s,t,\vec{c}',i) \neq V_{\mathcal{P}_5}^*(s,t,\vec{c}',i+1)$}
                    {
                        $\vec{x}^*(s,t,\vec{c}',i) \leftarrow \vec{x}^*\Big(\max \big\{s - \hat{\Upsilon}_{e_{i+1}}(t), 0 \big\}, t, \max \{ \vec{c}' - A_{:, i+1}, 0 \}, i+1 \Big)$\\
                        $x^*_{e_{i+1}}(s,t,\vec{c}',i) \leftarrow 1$ \tcp{Update}
                        \If{$\mathbf{A} \vec{x}^*(s,t,\vec{c}',i) \leq \vec{c}'$ is violated}
                        {
                            $V_{\mathcal{P}_5}^*(s,t,\vec{c}',i) \leftarrow V_{\mathcal{P}_5}^*(s,t,\vec{c}',i+1)$ \\
                            $\vec{x}^*(s,t,\vec{c}',i) \leftarrow \vec{x}^*(s,t,\vec{c}’,i+1)$ \\
                        }
                    }
                }
            }
            \tcc{Assign the solution of $i = 0$ to $\mathcal{P}_4(s,t)$}
            $\vec{x}^*_{\mathcal{P}_4}(s,t) \leftarrow \vec{x}^*(s,t,\vec{c},0)$ \\
        }
        \Return{$\big\{\vec{x}_{\mathcal{P}_4}^*(s,t)\big\}_{s\in\mathcal{S}(t)}$}
    \end{algorithm}
\end{figure}

The recursion starts from condition $s = 0$, $\vec{c} = \vec{0}$, and $i=|\mathcal{E}|$. \textbf{Algorithm \ref{algo2}} summarizes the main procedure. 
It is used to substitute Step 5 $\sim$ Step 7 of \textsc{Esdp}. Obviously, \textbf{Algorithm \ref{algo2}} is of 
$\mathcal{O}\big( |\mathcal{E}| \cdot |\mathcal{S}(t)| \cdot \prod_{k \in \mathcal{K}} c_k \big)$-complexity, i.e., 
$\{\mathcal{P}_4(s,t)\}_{s \in \mathcal{S}(t)}$ are solved in polynomial time. In the following content, we will show the relations between 
$\mathcal{P}_3(t)$ and $\{\mathcal{P}_4(s,t)\}_{s\in\mathcal{S}(t)}$, and analyze how the solution obtained by \textsc{Esdp} affects the regret 
$\textsc{Re}(T)$ defined in $\mathcal{P}_2$.

\subsection{Optimality and Regret Analysis}\label{s4.3}
In this section, we will analyze the upper bound of $\textsc{Re}(T)$ for \textsc{Esdp} when $T$ goes to infinity. The result is based on the relations 
between the optimal solutions of several problems we defined above. The problems and their optimal solutions are summarized in Tab. \ref{tab1} 
for quick reference. Our first result is that \textsc{Esdp} achieves the optimal statistical-based computation utility asymptotically with a certain probability.

\begin{theorem}
    \label{th1}
    (Probabilistic Asymptotical Optimality) By executing \textsc{Esdp} for problem $\mathcal{P}_3(t)$, 
    $\lim_{t \to \infty}\tilde{\textsc{U}}\big(\vec{x}_{\mathcal{P}_4}^*(t)\big)$ is at least
    \begin{equation}
        \max_{\vec{x}(t) \in \Omega(t)} \Bigg\{ \hat{\vec{\upsilon}}(t)^\mathrm{T} \vec{x}(t) + \sqrt{\hat{\vec{\sigma}}^2(t)^\mathrm{T}\vec{x}(t)} \Bigg\}
        \label{max1}
    \end{equation}
    with probability at most $\exp \bigg[ -\frac{1}{3} \Big( |\mathcal{L}| - \sum_{l \in \mathcal{L}} \rho_l(t) \Big)^2 \bigg].$
\end{theorem}
\begin{proof}
    Note that $\tilde{\textsc{U}}(\cdot)$ is the objective defined in $\mathcal{P}_3(t)$ and \eqref{max1} is exactly the optimal objective of $\mathcal{P}_3(t)$ without 
the approximate parameter $\delta(t)$. Before our proof, we define the set 
\begin{equation}
    \Phi(t) := \Big\{\vec{x}(t) \in \mathcal{X} \mid \eqref{cons1} \text{ holds at time }t \Big\}.
\end{equation}
Different from the set $\Omega(t)$, $\Phi(t)$ does not require constraint \eqref{cons2} to hold. Thus we have 
$\Omega(t) \subseteq \Phi(t)$. The following proof holds for every $t \in \mathcal{T}$.

\begin{table}[htbp]
    \begin{center}
    \caption{\label{tab1}Probelms and Their Optimal Solutions.}   
    \begin{tabular}{c|c}    
        \toprule
        {\textsc{Problems}} & {\textsc{Optimal Solutions}}\\[+0.1mm]
        \midrule
        $\mathcal{P}_1$ (defined over $\mathcal{T}$) & $\{\vec{x}^*(t)\}_{t \in \mathcal{T}}$ \\[+0.7mm]
        $\mathcal{P}_2$ (defined over $\mathcal{T}$) & $\{\vec{x}^*(t)\}_{t \in \mathcal{T}}$, because $\mathcal{P}_1 \equiv \mathcal{P}_2$ \\[+0.7mm]
        $\mathcal{P}_4(s,t)$ & $\vec{x}^*_{\mathcal{P}_4}(s,t)$, also $\vec{x}^*(s,t,\vec{c},0)$ \\[+0.7mm]
        $\max\{\mathcal{P}_4(s,t)\}_{s \in \mathcal{S}(t)}$ & $\vec{x}^*_{\mathcal{P}_4}(t)$ (by Step 8 of \textsc{Esdp}) \\[+0.7mm]
        $\mathcal{P}_5(s,t,\vec{c},i)$ & $\vec{x}^*(s,t,\vec{c},i)$ (by \textbf{Algorithm \ref{algo2}}) \\[+0.7mm]
        \bottomrule   
    \end{tabular}  
    \end{center}
\end{table}

By the definitions \eqref{st1}, \eqref{scal-mean}, \eqref{scal-var} and the fact $\tilde{\vec{Z}} \in [0,1]^{|\mathcal{E}|}$, we have 
\begin{flalign}
    \hat{\vec{\upsilon}}(t) \leq \frac{\hat{\vec{\Upsilon}}(t)}{\xi(t)} \leq \frac{1}{\xi(t)} \vec{1} + \hat{\vec{\upsilon}}(t).
    \label{23}
\end{flalign}
Thus, 
\begin{flalign}
    &\max_{\vec{x}(t) \in \Omega(t)} \hat{\vec{\upsilon}}(t)^\mathrm{T} \vec{x}(t) + \sqrt{\hat{\vec{\sigma}}^2(t)^\mathrm{T} \vec{x}(t)} \nonumber \\
    &\leq \frac{1}{\xi(t)} \max_{\vec{x}(t) \in \Omega(t)} \hat{\vec{\Upsilon}}(t)^\mathrm{T} \vec{x}(t) + \sqrt{\hat{\vec{\Sigma}}^2(t)^\mathrm{T} \vec{x}(t)}.
    \label{proof1}
\end{flalign} 
Further, the RHS of \eqref{proof1} staisfies
\begin{flalign}
    &\max_{\vec{x}(t) \in \Omega(t)} \hat{\vec{\Upsilon}}(t)^\mathrm{T} \vec{x}(t) + \sqrt{\hat{\vec{\Sigma}}^2(t)^\mathrm{T} \vec{x}(t)} \nonumber \\
    &= \max_{s \in \mathcal{S}(t)} \max_{\hat{\vec{\Upsilon}}(t)(t)^\mathrm{T} \vec{x}(t) = s, \vec{x}(t) \in \Omega(t)} \Bigg\{s + \sqrt{\hat{\vec{\Sigma}}^2(t)^\mathrm{T} \vec{x}(t)} \Bigg\} \nonumber \\
    &\leq \max_{s \in \mathcal{S}(t)} \max_{\hat{\vec{\Upsilon}}(t)^\mathrm{T} \vec{x}(t) \geq s, \vec{x} \in \Omega(t)} \Bigg\{s + \sqrt{\hat{\vec{\Sigma}}^2(t)^\mathrm{T} \vec{x}(t)} \Bigg\} \nonumber \\
    &\leq \max_{s \in \mathcal{S}(t)} \max_{\hat{\vec{\Upsilon}}(t)^\mathrm{T} \vec{x}(t) \geq s, \vec{x} \in \Phi(t)} \Bigg\{s + \sqrt{\hat{\vec{\Sigma}}^2(t)^\mathrm{T} \vec{x}(t)} \Bigg\}.
    \label{25}
\end{flalign}
The RHS of \eqref{25} is exactly $\max \{\mathcal{P}_4(s,t)\}_{s \in \mathcal{S}(t)}$. The upper bound of it should be 
$s^\star + \sqrt{\hat{\vec{\Sigma}}^2(t)^\mathrm{T} \vec{x}_{\mathcal{P}_4}^*(t)}$ if no channel is shut down forcibly, i.e., Step 10 $\sim$ Step 16 are not executed by \textsc{Esdp}. 
To quantify the probability that no channels are forcibly shut down, we use the result of Chernoff Bounds. The upper tail of Chernoff Bounds is stated as follows.

\textit{If $X_1, ..., X_n \in \{0,1\}$ are mutually independent, then $\forall x \geq \mu$, where $\mu := \mathbb{E}\big[ \sum_{i}X_i \big]$, we have
\begin{equation*}
    \Pr\bigg[\sum_{i}X_i \geq x\bigg] \leq e^{x - \mu}\bigg(\frac{\mu}{x}\bigg)^x.
\end{equation*}}Based on this conclusion, we can further derive that 
\begin{equation}
    \Pr\bigg[ \sum_i X_i \geq (1 + \varepsilon) \mu \bigg] \leq \bigg( \frac{e^\varepsilon}{(1+\varepsilon)^{1+\varepsilon}} \bigg)^\mu,
    \label{chernoff2}
\end{equation}
where $\varepsilon \geq 0$. 

{\color{black}
With the Taylor-series expansion for $\ln(x + 1)$ at $x = 0$, we have
\begin{equation*}
    \ln (1 + \varepsilon) = \sum_{n=1}^\infty \frac{(-1)^{n+1} \varepsilon^n}{n} = \varepsilon - \frac{\varepsilon^2}{2} + \frac{\varepsilon^3}{3} - ... \geq  \varepsilon - \frac{\varepsilon^2}{2}.
\end{equation*}
Thus, we have
\begin{equation*}
    \frac{1}{\ln (1 + \varepsilon)} \leq \frac{1}{\varepsilon(1 - \frac{1}{2}\varepsilon)} = \frac{1}{\varepsilon} + \frac{1}{2-\varepsilon} \leq  \frac{1}{\varepsilon} + \frac{1}{2}.
\end{equation*}
}
Applying the inequality to the RHS of \eqref{chernoff2}, we can get
\begin{equation}
    \Pr\bigg[ \sum_i X_i \geq (1 + \varepsilon) \mu \bigg] \leq \exp \bigg( - \frac{\varepsilon^2 \mu}{3} \bigg).
    \label{chernoff3}
\end{equation}
Replacing $X_i$ with $\mathds{1}_l$ and $(1 + \varepsilon) \mu$ with $|\mathcal{L}|$, \eqref{chernoff3} is transformed into 
\begin{equation}
    \Pr\bigg[ \sum_{l \in \mathcal{L}} \mathds{1}_l = |\mathcal{L}| \bigg] \leq \exp \bigg[ -\frac{1}{3} \Big( |\mathcal{L}| - \sum_{l \in \mathcal{L}} \rho_l(t) \Big)^2 \bigg],
    \label{pr}
\end{equation}
which exactly quantifies the probability that every port yields at least one job. In this case, no channel $(l,r) \in \mathcal{E}$ is shut down forcibly. 
Thus, with this probability, the RHS of \eqref{25} staisfies
\begin{flalign}
    &\qquad \max_{s(t) \in \mathcal{S}} \max_{\hat{\vec{\Upsilon}}(t)^\mathrm{T} \vec{x}(t) \geq s, \vec{x}(t) \in \Phi(t)} \Bigg\{s + \sqrt{\hat{\vec{\Sigma}}^2(t)^\mathrm{T} \vec{x}(t)} \Bigg\} \nonumber \\
    &= s^\star + \sqrt{\hat{\vec{\Sigma}}^2(t)^\mathrm{T} \vec{x}_{\mathcal{P}_4}^*(t)} \quad \triangleright s^\star \text{\textit{ staisfies \eqref{P4} with prob. }} \text{\eqref{pr}} \nonumber \\
    &\leq \hat{\vec{\Upsilon}}(t)^\mathrm{T} \vec{x}^*_{\mathcal{P}_4}(t) + \sqrt{\hat{\vec{\Sigma}}^2(t)^\mathrm{T} \vec{x}_{\mathcal{P}_4}^*(t)} \qquad \triangleright \text{\eqref{s}} \nonumber \\
    &\leq \Big(\vec{1} + \xi(t) \hat{\vec{\upsilon}}(t)\Big)^\mathrm{T} \vec{x}^*_{\mathcal{P}_4}(t) + \sqrt{\hat{\vec{\Sigma}}^2(t)^\mathrm{T} \vec{x}_{\mathcal{P}_4}^*(t)} \qquad \triangleright \text{\eqref{23}} \nonumber \\
    &\leq \xi(t) \bigg( \delta(t) + \hat{\vec{\upsilon}}(t)^\mathrm{T} \vec{x}^*_{\mathcal{P}_4}(t) + \sqrt{\hat{\vec{\sigma}}^2(t)^\mathrm{T}\vec{x}_{\mathcal{P}_4}^*(t)} \bigg).
    \label{final-proof}
\end{flalign}
Combining the result of \eqref{proof1}, \eqref{25}, and \eqref{final-proof}, the following inequality holds for all the time $t \in \mathcal{T}$:
\begin{flalign}
    \max_{\vec{x}(t) \in \Omega(t)} \Bigg\{ \hat{\vec{\upsilon}}(t)^\mathrm{T} \vec{x}(t) + \sqrt{\hat{\vec{\sigma}}^2(t)^\mathrm{T}\vec{x}(t)} \Bigg\}
    \leq \tilde{\textsc{U}}\big(\vec{x}^*_{\mathcal{P}_4}(t)\big).
\end{flalign} 
When $t \to \infty$ and $\delta(t) \to 0$, the result is tightly bounded.
\end{proof}
The theorem shows that \textsc{Esdp} can achieve approximately 
optimal statistical-based computation utility at each time slot with a certain probability. This optimality is important for minimizing the regret because 
it builds the upper bound of the optiaml computation utility $\tilde{\vec{\upsilon}}^\mathrm{T} \vec{x}^*(t)$ at each time $t$. The probabilistic regret 
upper bound is given by the following theorem.

\begin{theorem}
    \label{th2}
    (Regret Upper Bound under Certain Conditions) By executing the \textsc{Esdp} algorithm, as $T \to \infty$, $\textsc{Re}(T)$ is upper bounded by
    \begin{flalign}
         \mathcal{O} \bigg( \ln T \cdot \frac{ |\mathcal{E}| \cdot \big( \ln \vec{x}^* \big)^2 }{\min_{t \in \mathcal{T}} \Delta\big(\vec{x}_{\mathcal{P}_4}^*(t)\big)} \bigg)
         \label{bound}
    \end{flalign}
    with probability at most $\exp \Big( -\frac{1}{3} \big( |\mathcal{L}| - \sum_{l \in \mathcal{L}} \rho_l(t) \big)^2 \Big)$.
    In \eqref{bound}, $\Delta\big(\vec{x}_{\mathcal{P}_4}^*(t)\big)$ is introduced by \eqref{gap}, and $\vec{x}^*$ is defined as 
    \begin{equation}
        \vec{x}^* := \argmax_{\forall t \in \mathcal{T}: \vec{x}^*(t)} \| \vec{x}^*(t)\|_1.
    \end{equation}
\end{theorem}
\begin{proof}
    \color{black}
    The result is immediate with \textbf{Theorem \ref{th1}} and \textbf{Theorem 4.4} of \cite{semi-bandit}. 
    The technique is to define three events 
    $\mathcal{A}(t)$, $\mathcal{B}(t)$, $\mathcal{C}(t)$ at each time $t$: 
    \begin{flalign}
        \left\{
        \begin{array}{l}
        \mathcal{A}(t) := \big\{ \big|\big(\tilde{\vec{\upsilon}} - \hat{\vec{\upsilon}}(t)\big)^\mathrm{T} \vec{x}^*(t)\big| \geq \sqrt{\hat{\vec{\sigma}}^2(t)^\mathrm{T} \vec{x}^*(t)} \big\} \\
        \mathcal{B}(t) := \big\{ \Delta(\vec{x}_{\mathcal{P}_4}^*(t)) \leq 4 \delta(t) \big\} \\
        \mathcal{C}(t) := \overline{\mathcal{A}(t)} \cup \overline{\mathcal{B}(t)},
        \end{array}
        \right.
    \end{flalign}
    and study the sum of the upper bound of $\textsc{Re}(T)$ under these events respectively. Which of these events will happen depends on the accuracy 
    of the estimations $\hat{\vec{\upsilon}}(t)$. Considering that the proof is similar to the proof presented in \cite{semi-bandit}, we will not 
    demonstrate the complete proof here. 
\end{proof}

{\color{black}
\subsection{Extending to Gang Scheduling}
\textsc{Esdp} can be extended to the Gang scheduling scenarios, where the scheduling decisions for the task instances of a 
job follows the \textsc{All-or-Nothing} property. In other words, only when \textit{all} tasks\footnote{In practice, not all tasks 
of a job need to be scheduled. In Kubernetes, the job submitter can specify the minimum number of tasks that 
must be scheduled successfully. In the following, we use $m_l(t)$ to represent the minimum number of tasks that should be 
scheduled at time $t$ of the type-$l$ job.} of a job are successfully scheduled, the job could be launched. Gang 
scheduling is required for multi-server jobs such as distributed DNN trainings and Message Passing Interface (MPI) jobs. 
Take the DNN training with the parameter server (PS)-worker architecture as an exmaple, at least one PS and one worker 
are successfully scheduled, the training could start.

In the following, we show briefly how Gang Scheduling can be modeled. Let us re-define $\vec{x}(t)$ as 
\begin{equation*}
    \vec{x}(t) := \Big[ x_{(q,r)}^l(t) \Big]_{q \in \mathcal{Q}_l, r \in \mathcal{R}_l, l \in \mathcal{L}},
\end{equation*}
where $\mathcal{Q}_l$ stores the indices of tasks for the type-$l$ job. Then, we have the following new constraints:
\begin{equation*}
    \left\{
        \begin{array}{ll}
            \sum_{r \in \mathcal{R}_l} \sum_{q \in \mathcal{Q}_l} x_{(q,r)}^l (t) \geq m_l (t) & \forall l, t \\
            \sum_{l \in \mathcal{L}} \sum_{q \in \mathcal{Q}_l} a_{(q,k)}^l x_{(q,r)}^l(t) \leq c_{(k,r)} &\forall k, r, t,
        \end{array}
    \right.
    \label{new-cons}
\end{equation*}
where $m_l(t)$ is the minimum number of tasks to be executed, $a_{(q,k)}^l$ is the requirement of the type-$k$ 
resource for the $q$-th task of the type-$l$ job, and $c_{(k,r)} \in \mathbb{N}^+$ is the number of the type-$k$ computing devices 
available to server $r$. The same to \eqref{cons1}, the new constraint also has the form of $\mathbf{A}\vec{x} \leq \vec{c}$. The new problem 
can be solved by a similar approach to \textsc{Esdp} after several mathematical transformations.
}
% ======================================================================================================================================================

\section{Numerical Results}\label{s5}
In this section, we conduct extensive simulations to validate the performance of \textsc{Esdp}.  
We firstly verify the performance of \textsc{Esdp} against several handcrafted benchmarking policies 
on the \textsc{Aou}. Then, we analyze the generality and robustness of it under different cluster settings.
The simulations are conducted on a server with 48 Intel Xeon Silver 4214 CPUs, 256 GB memory, and 2 Tesla P40 GPUs. 

{\color{black}\textit{Traces.} We use the data from cluster-trace-v2018 
of the Alibaba Cluster Trace Program\footnote{https://github.com/alibaba/clusterdata} to generate our experiment 
experiments. Specifically, we leverage the specifications of the machines, the arrival patterns and resource requirements 
of different kind of jobs to set resource capacities $\{ c_k \}_{k \in \mathcal{K}}$, job arrival probabilities 
$\{ \rho_l \}_{l \in \mathcal{L}}$, and device requirements of jobs $\{ \vec{a}_k \}_{k \in \mathcal{K}}$.

\textit{Default Scenario Settings}. In default settings, our simulation environment has 40 servers, each equipped with 3 types 
of computing devices (CPUs, MEM, and GPUs), and 8 multi-sever job types of different resource requirements. Job arrival 
probabilities $\{ \rho_l \}_{l \in \mathcal{L}}$ are setted to adjust the job arrival status with Bernoulli Distributions. 
$\{ \rho_l \}_{l \in \mathcal{L}}$ are applied on the basis of the actual arrival patterns from the trace to increase stochasticity. 
Although $\{ \vec{a}_k \}_{k \in \mathcal{K}}$ are retrieved from the trace data, we still need to set the equipped resource 
limits to eliminate inappropriate settings which could lead to the solution space of problem $\mathcal{P}_1$ being null. 
Specifically, we denote by $\overline{\| \mathbf{A} \|_2}$ and $\underline{\| \mathbf{A} \|_2}$ the upper bound and the lower bound 
of $\{A_{ij}\}_{\forall i, j}$, and set them to $2$ and $1$ in default, respectively. Correspondingly, we use 
$\overline{\| \vec{c} \|_2}$ and $\underline{\| \vec{c} \|_2}$ to represent the upper bound and the lower bound of $\vec{c}$, and set 
them to $2$ and $1$ in default, respectively. The settings of these bounds are normalized. For each computation utility 
$v_{(l,r)}$, $(l,r) \in \mathcal{E}$, we generate it from a Normal distribution as follows:
\begin{equation*}
    \mathcal{N} \bigg(\mu_{(l,r)} \sim U(0.1, 1), \sigma_{(l,r)} = \frac{\mu_{(l,r)}}{2} \bigg).
\end{equation*}
Correspondingly, for each device type $k \in \mathcal{K}$, the operating cost $f_k (a_k^{(l,r)})$ is generated from the Normal distribution 
$\mathcal{N} (0.5, 0.1)$. Note that the settings we adopt are only required to make the stochastic problem $\mathcal{P}_1$ feasible. 
\textsc{Esdp} is robust enough to make scheduling decisions of high system efficiency. The robustness will be demonstrated in detail in the following 
content. Besides, note that \textsc{Esdp} has no assumptions 
on the distributions of the valuations $\{v_{(l,r)}\}_{(l,r) \in \mathcal{E}}$. The Normal distributions we used here are only for problem 
construction. The default time slot length is $2000$.
}

\begin{figure*}[htbp]
    \centering
    \color{black}
    \begin{minipage}[t]{0.3\textwidth}
        \centering
        \includegraphics[width=1\textwidth]{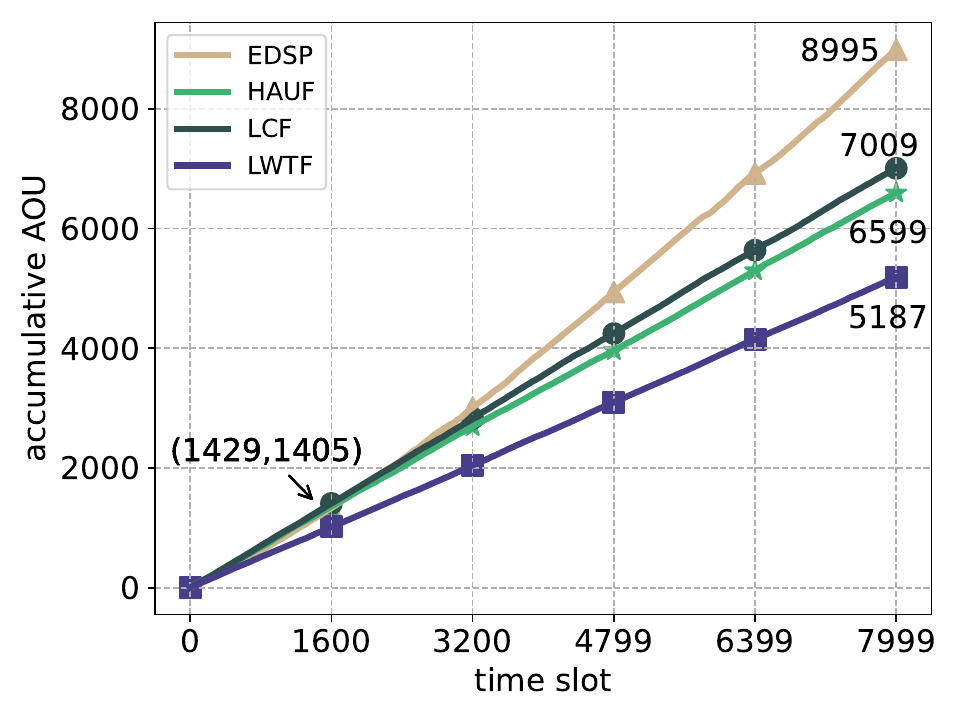}
		\caption{The \textsc{Aou} vs. time slots.}
		\label{super}
    \end{minipage}
    \begin{minipage}[t]{0.3\textwidth}
        \centering
        \includegraphics[width=1\textwidth]{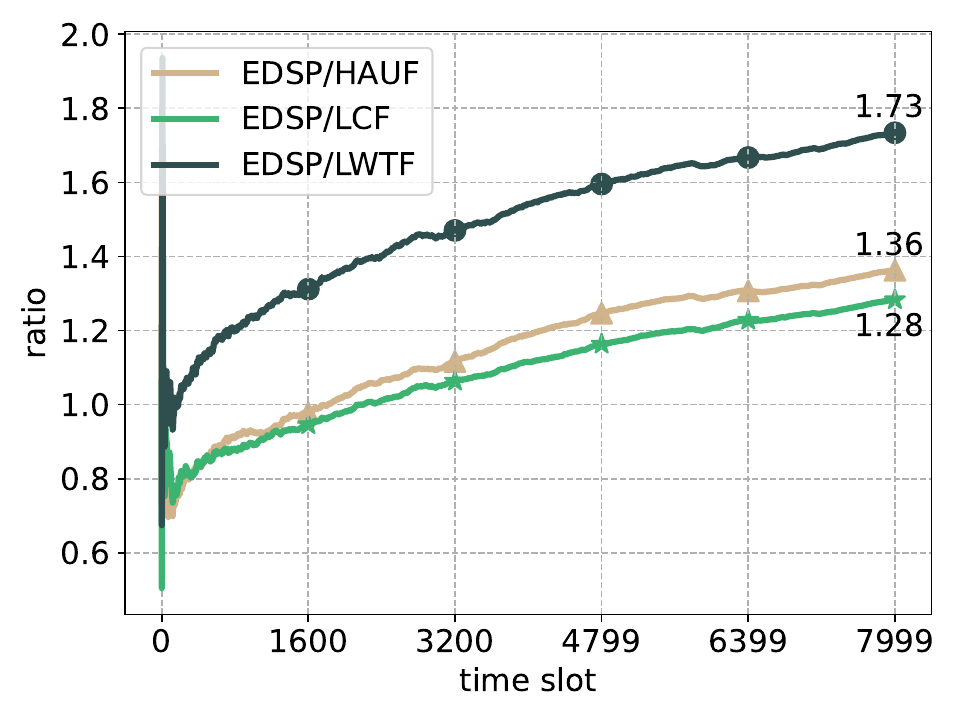}
        \caption{The ratio between the \textsc{Aous}.}
        \label{super2}
    \end{minipage}
    \begin{minipage}[t]{0.297\textwidth}
        \centering
        \includegraphics[width=1\textwidth]{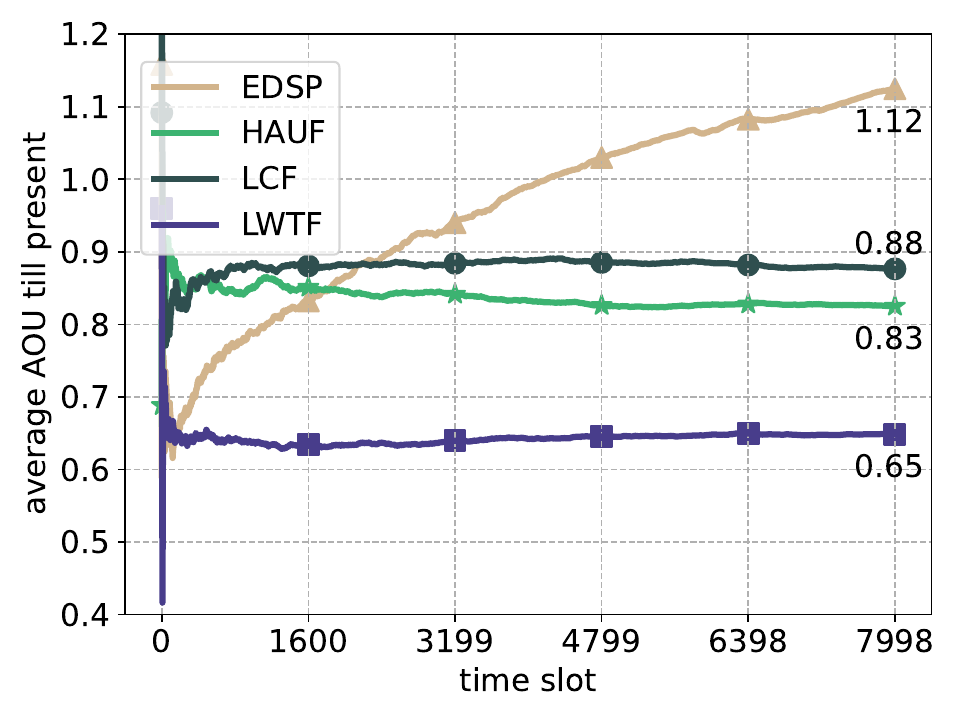}
        \caption{The average \textsc{Aou} vs. time slots.}
        \label{astring}
    \end{minipage}
\end{figure*}

\textit{Default Algorithmic Settings.} When we implement \textsc{Esdp}, $\max_{t' \in \mathcal{T}} \big\{\max_{\vec{x} \in \Omega(t')} \| \vec{x} \|_1 \big\}$ 
is calculated as $\alpha |\mathcal{E}|$, 
where $\alpha \in [0, 1]$ is a coefficient by default to be 0.5. We set $\delta(t)$ and $g(t)$ as $\big(\ln(\ln(t+1)+1)+1\big)^{-1}$ and 
$\ln(t+1)+4\ln\big(\ln(t+1)+1\big) \cdot \alpha |\mathcal{E}|$, respectively in default. Considering that those two sequences significantly 
affect the effectiveness of \textsc{Esdp}, we will comprehensively discuss their variations in Sec. \ref{s5.3}.

\begin{table}[htbp]
    \begin{center}
    \caption{\label{tab2} Default parameter settings.} 
    \begin{tabular}{c|c|c|c}    
        \toprule
        {\textsc{Param.}} & {\textsc{Value}} & {\textsc{Param.}} & {\textsc{Value}}\\[+0.1mm]
        \midrule
        $| \mathcal{L}| $ & $8$ & $| \mathcal{R} |$ & $40$ \\[+0.7mm]
        $\overline{\| \mathbf{A} \|_2}$ & $2$ & $\underline{\| \mathbf{A} \|_2}$ & $1$ \\[+0.7mm]
        $\alpha$ & $0.5$ & $\{\rho_l\}_{l \in \mathcal{L}}$ & $0.9$\\[+0.7mm]
        $\overline{\| \vec{c} \|_2}$ & $2$ & $\underline{\| \vec{c} \|_2}$ & $1$ \\[+0.7mm]
        $K$ & $3$ & $T$ & $2000$ \\[+0.7mm]
        $\{ f_k \}_{k \in \mathcal{K}}$ & $\sim \mathcal{N} (0.5, 0.1)$ & $\{ \rho_{(l,r)} \}_{(l,r) \in \mathcal{E}}$ & $0.1$ \\[+0.7mm]
        \bottomrule   
    \end{tabular}
    \end{center}
\end{table}

\textit{Baselines}. \textsc{Esdp} is compared with the following handcrafted baselines.

\begin{itemize}
    \item \textit{\color{black}The Accumulative Utility First (HAUF)}: HSWF is different from \textsc{Esdp} in the following ways. At each time $t$, 
    $\vec{Z}(t)$ is estimated as the average of historical observations. With the estimate, HSWF ranks each port in the 
    descending order of $\sum_{r \in \mathcal{R}_l} x_{(l,r)} (t) \big( Z_{(l,r)}(t) - \sum_{k \in \mathcal{K}} f_k(a_k^{(l,r)}) \big)$, 
    and set the corresponding $x_{(l,r)}(t)$ as 1 in turn until \eqref{cons1} can not be satisfied.
    \item \textit{The Lowest Cost First (LCF)}: Similar to HSWF, at each time $t$, $\vec{Z}(t)$ is estimated as the average of 
    historical observations. Then, LCF ranks each job (non-empty port) in the ascending order of cost $\sum_{k \in \mathcal{K}} f_k(a_k^{(l,r)})$, 
    and set the corresponding $x_{(l,r)}(t)$ as 1 in turn until \eqref{cons1} can not be satisfied.
    \item \textit{The Longest Waiting Time First (LWTF)}: LWTF is different from \textsc{Esdp} in two ways. Firstly, $\vec{Z}(t)$ is estimated as 
    the average of historical observations. Secondly, LWTF ranks each port in the descending order of the waiting time of jobs yield from that port, 
    and set the corresponding $x_{(l,r)}(t)$ as 1 in turn until \eqref{cons1} can not be satisfied.
\end{itemize}

Note that we do not implement heuristics such as the Genetic Algorithm for comparison. This is because $\mathcal{P}_1$ 
is a stochastic optimization problem and traditional heuristics need to be revised carefully to match it. All of the three baselines use 
a similar method to estimate the historical valuation of each channel. With the estimate, the stochastic optimization problem is 
transformed into a deterministic one. Essentially, heuristics can be implemented by following a similar approach. However, a big problem 
that cannot be ignored is that heuristics are time-consuming with a non-polynomial complexity. Heuristics have to be called in every 
time slot, which could be very slow when the time slot length is large. 

\subsection{Performance Verification}\label{s5.2}

In the first part, we demonstrate how the average \textsc{Aou} changes as time slot increases. As Fig. \ref{super} shows, 
\textsc{Esdp} outperforms the baselines by up to nearly 73\%, 36\%, and 28\%, respectively within 8000 time slots. 
In the beginning, HSWF performs better than EDSP, but as the time slots increase, \textsc{Esdp} gradually outperforms 
HSWF, and the gap between them keeps widening. \textsc{Esdp} is able to surpass HSWF because that, unlike HSWF, which 
does not adjust its strategy, EDSP constantly updates its strategy with the explorated valuation distributions. Besides, 
we also demonstrate the ratio between the \textsc{Aou} achieved by \textsc{Esdp} and the baselines in Fig. \ref{super2}. 
We can conclude that, the performance of \textsc{Esdp} increases significantly when the time slots available to explore 
increase. The reason lies in that more time slots leads to more approximate estimate to $\{v_{(l,r)}\}_{(l,r) \in \mathcal{E}}$.

\begin{figure*}[htbp]
    \centering
    \color{black}
    \begin{minipage}[t]{0.26\textwidth}
        \centering
        \includegraphics[width=1\textwidth]{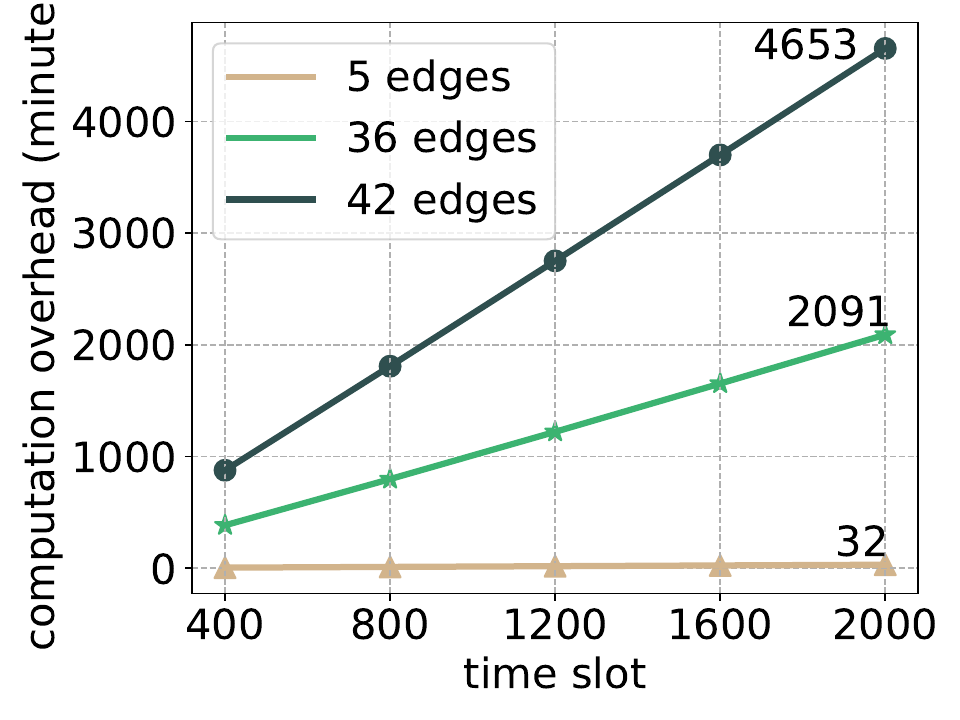}
        \caption{Computation overheads.}
        \label{TimeOver}
    \end{minipage}
    \begin{minipage}[t]{0.26\textwidth}
            \centering
            \includegraphics[width=1\textwidth]{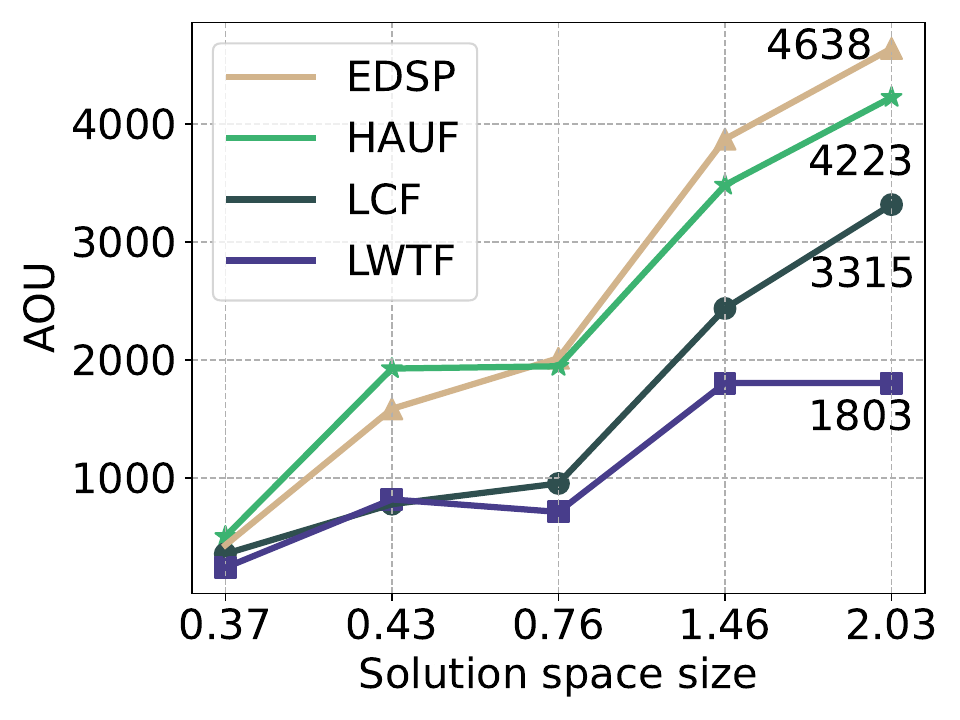}
            \caption{\textsc{Aou} vs. $\mathcal{X}$. }
            \label{range}
    \end{minipage}
    \begin{minipage}[t]{0.26\textwidth}
            \centering
            \includegraphics[width=1\textwidth]{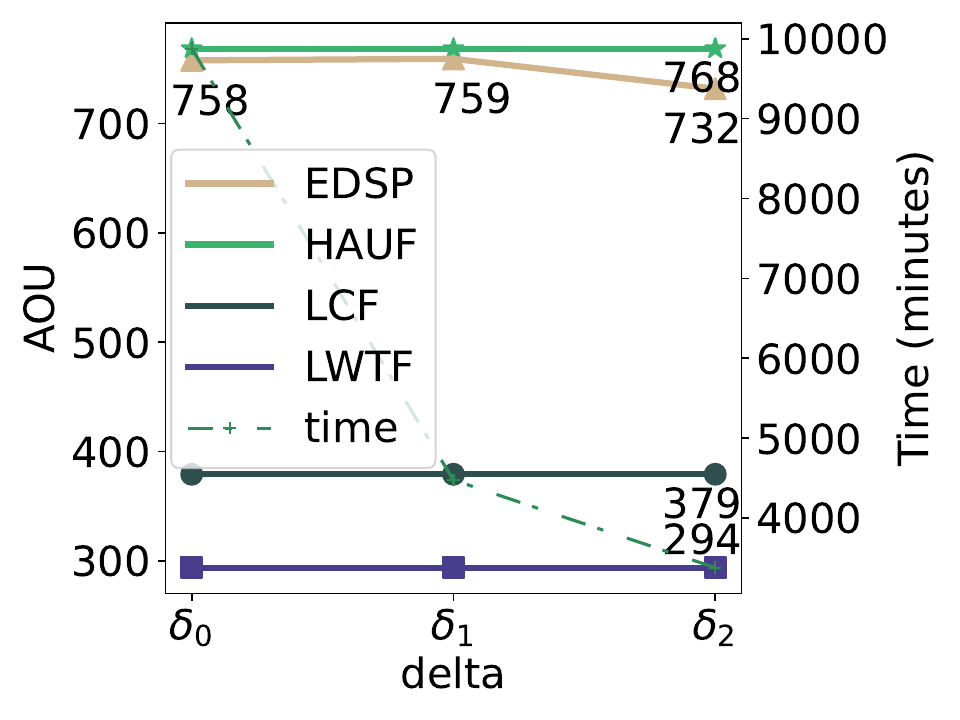}
            \caption{\textsc{Aou} vs. $\{ \delta(t) \}_{t \in \mathcal{T}}$.}
            \label{delta}
    \end{minipage}
\end{figure*}

In Fig. \ref{astring}, we calculate the average \textsc{Aou} in this way: for each time slot length $T$, the $y$-axis value 
is $\frac{1}{T} \sum_{t=1}^T \textsc{U}(\vec{x}(t))$. Different from the baselines, the average \textsc{Aou} of 
\textsc{Esdp} increases steep and later flattens, which verifies that the \textsc{Aou} converges to an underlying upper bound 
(the \textsc{Aou} achieved by the oracle). The computation overhead of \textsc{Esdp} under different scales of the bipartite 
graph is shown in Fig. \ref{TimeOver}. {\color{black}It is interesting to find that the rewards oscillate at the beginning time slots. 
One of the leading factors is that \textsc{Esdp} is boosted with a well designed initial solution. No surprisingly, the rewards 
achieved in the beginning can be easily surpassed when the time slot is sufficiently large. }

\begin{figure*}[htbp]
    \centering
    \color{black}
    \begin{minipage}[t]{0.26\textwidth}
        \centering
        \includegraphics[width=1\textwidth]{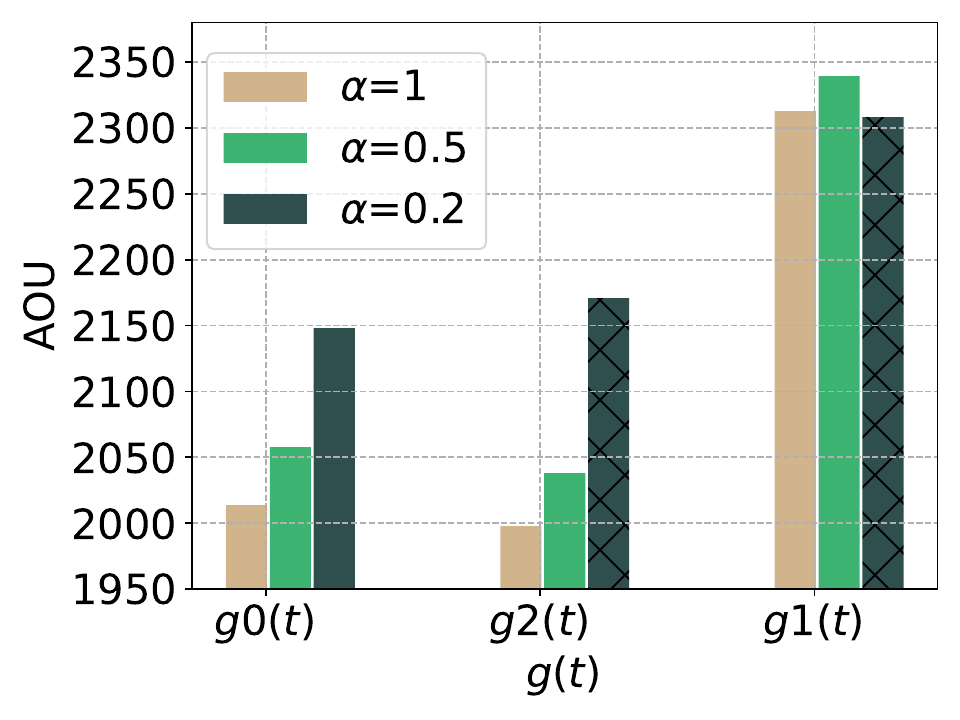}
        \caption{\textsc{Aou} vs. $\{ g(t) \}_{t \in \mathcal{T}}$.}
        \label{gt}
    \end{minipage}
    \begin{minipage}[t]{0.26\textwidth}
            \centering
            \includegraphics[width=1\textwidth]{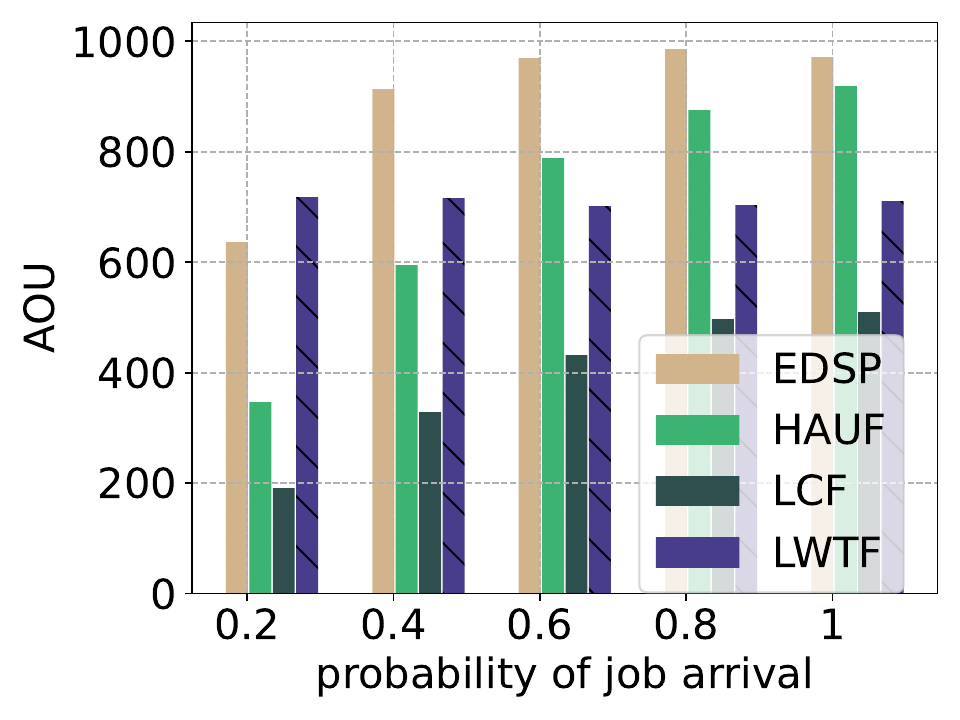}
            \caption{\textsc{Aou} vs. $\rho$.}
            \label{rho}
    \end{minipage}
    \begin{minipage}[t]{0.26\textwidth}
            \centering
            \includegraphics[width=1\textwidth]{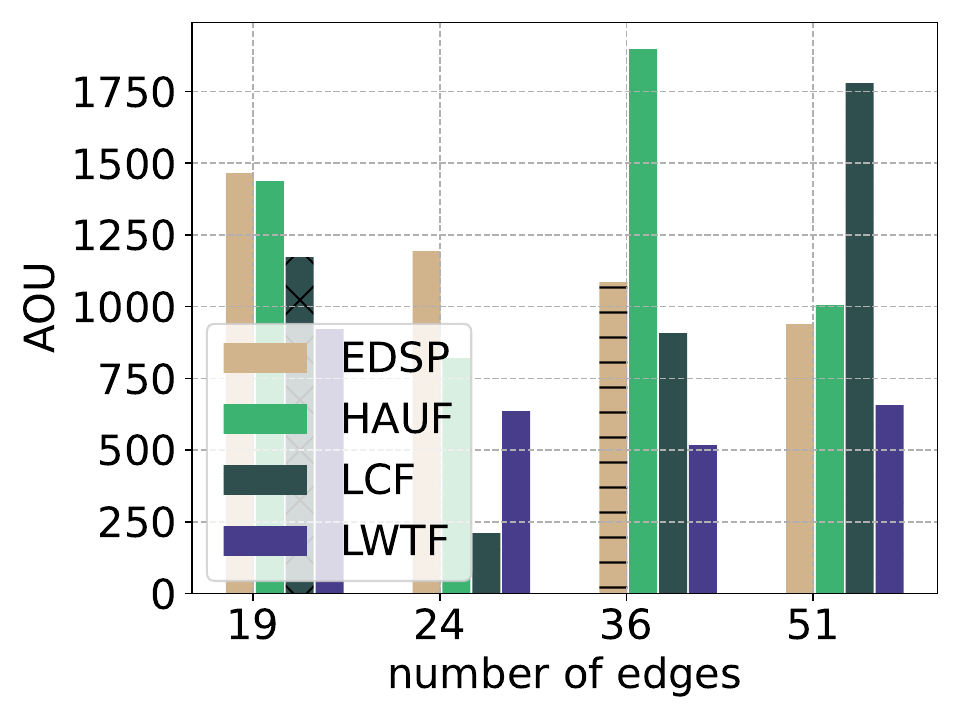}
            \caption{\textsc{Aou} vs. $|\mathcal{E}|$.}
            \label{xDim}
    \end{minipage}
\end{figure*}

\subsection{Sensitivity Analysis}\label{s5.3}
{\color{black}In this section, we give a brief analysis on several important parameter settings. The first problematic parameter we test 
is the size of the solution space $\mathcal{X}$, which is tuned by $\mathbf{A}$ and $\vec{c}$. Recall that 
$\mathbf{A} := \{\vec{a}_k\}_{k \in \mathcal{K}}$ and $\vec{c} := \{ c_k  \}_{k \in \mathcal{K}}$ are respectively the device 
requirements of each type of jobs and the device capacities of the cluster. The $x$-axis of Fig. \ref{range} is 
$\| \mathbf{A}^{-1} \vec{x} \|_2$. Without doubt, the \textsc{Aou} increases with the growth of $\mathcal{X}$ for all the 
algorithms because the number of can-be-processed jobs increase. Even though, \textsc{Esdp} has the highest growth in the 
\textsc{Aou} because it can fully exploit the estimated valuations. }

The first algorithmic parameter we pay attention to is the sequence $\{\delta_t\}_{t \in \mathcal{T}}$, which is 
used to relax the NP-hard problem $\mathcal{P}_2$ to a polynomial one. The three $\{\delta_t\}_{t \in \mathcal{T}}$ 
shown in Fig. \ref{delta} are $\big( \ln (t + 1) + 1 \big)^{-1}$, $\big( \ln (\ln (t + 1) + 1) \big)^{-1}$, and 
$\big( \ln ((\ln (\ln (t + 1) + 1)) + 1) + 1 \big)^{-1}$, respectively. Fig. \ref{delta} demonstrates that different settings of 
the sequence has little effect on the performance of \textsc{Esdp}, but strong affect on the computation overhead. 
Another algorithmic parameter we are interested on is $\{g(t)\}_{t \in \mathcal{T}}$, which is used to estimate 
the variance $\eqref{st2}$. The three $\{g(t)\}_{t \in \mathcal{T}}$ demonstrated in Fig. \ref{gt} are 
$\ln(t+1)+4 \ln(\ln(t+1)+1) \cdot \alpha |\mathcal{E}|$, $4 \ln(\ln(t+1)+1)\cdot \alpha |\mathcal{E}|$, and $\ln(t+1)$, respectively. 
We can find that the third setting has an overwhelming advantage. The reason is that, theoretically, $g(t)$ acts as a 
balancer between exploration and exploitation. A smaller $g(t)$ leads to a higher tendency to exploitation.

Fig. \ref{rho} and Fig. \ref{xDim} demonstrate the impact of job arrival rate $\rho$ and the density of the bipartite graph. 
From Fig. \ref{rho} we can find that, with the increase of $\rho$, the \textsc{Aou} achieved by nearly all the algorithms 
also goes up. The result is obvious because more jobs can be processed within service capacities when $\rho$ increases. 
{\color{black}It is interesting to find that, increasing the job arrival probability can lead to a high resource utilization, 
thereby increasing the \textsc{Aou}. However, a large job arrival probability also brings in a fierce resource contention. 
A direct consequence of it is that, for \textsc{Esdp}, many elements in the vector $\vec{x}(t)$ fall into the interior of $\mathcal{X}$, 
rather than the boundaries, thereby leading to a reward reduction. The phenomenon can be observed when moving $\rho$ from $0.8$ to $1.0$.}
Fig. \ref{xDim} demonstrates the impact of the service locality constraint. When the number of edges increases in the bipartite graph, 
which means the service locality constraint is relaxed, the solution space $\mathcal{X}$ becomes larger. It significantly increases 
the difficulty of searching the optimal solution for \textsc{Esdp}.

\subsection{\color{black}Scalability Analysis}\label{s5.4}
\begin{figure*}[htbp]
    \centering
    \color{black}
    \begin{minipage}[t]{0.26\textwidth}
        \centering
        \includegraphics[width=1\textwidth]{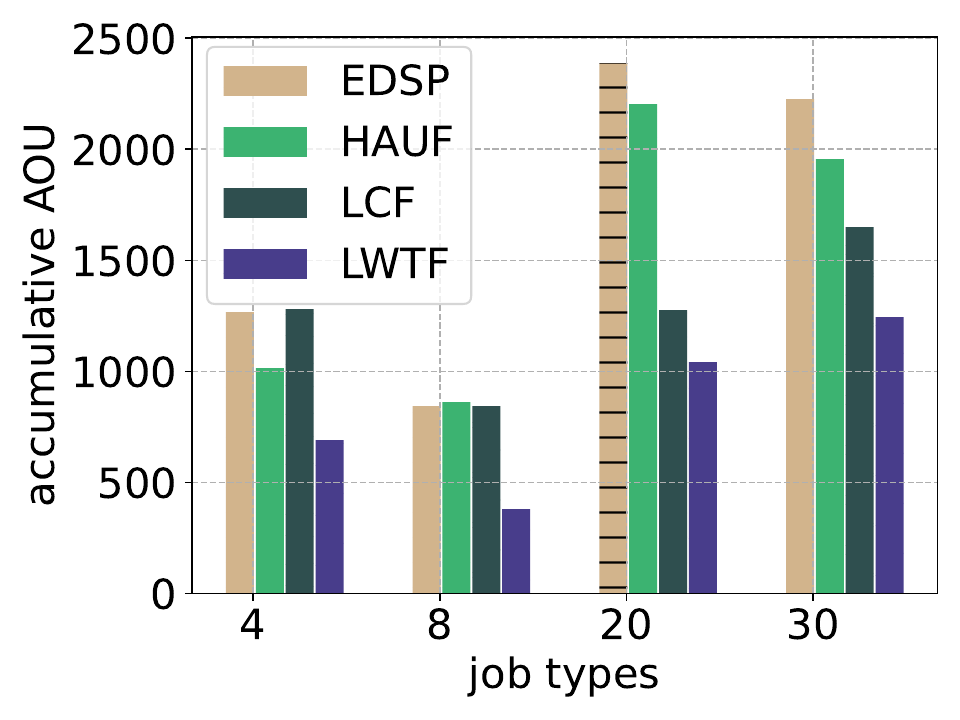}
        \caption{\textsc{Aou} vs. $|\mathcal{L}|$.}
        \label{L}
    \end{minipage}
    \begin{minipage}[t]{0.26\textwidth}
        \centering
        \includegraphics[width=1\textwidth]{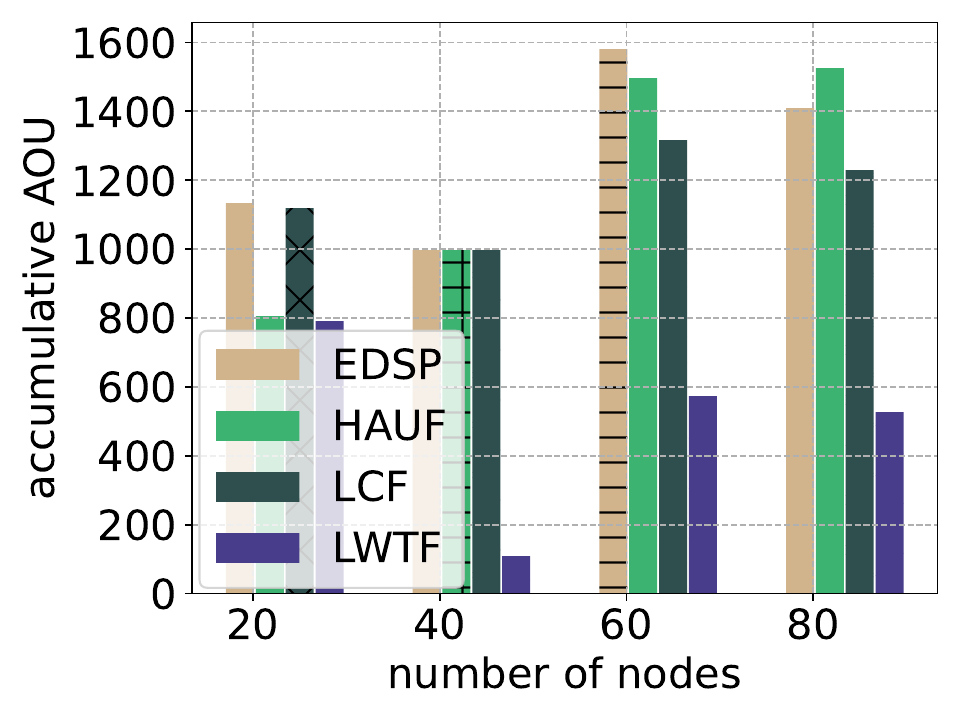}
        \caption{\textsc{Aou} vs. $|\mathcal{R}|$.}
        \label{R}
    \end{minipage}
    \begin{minipage}[t]{0.26\textwidth}
            \centering
            \includegraphics[width=1\textwidth]{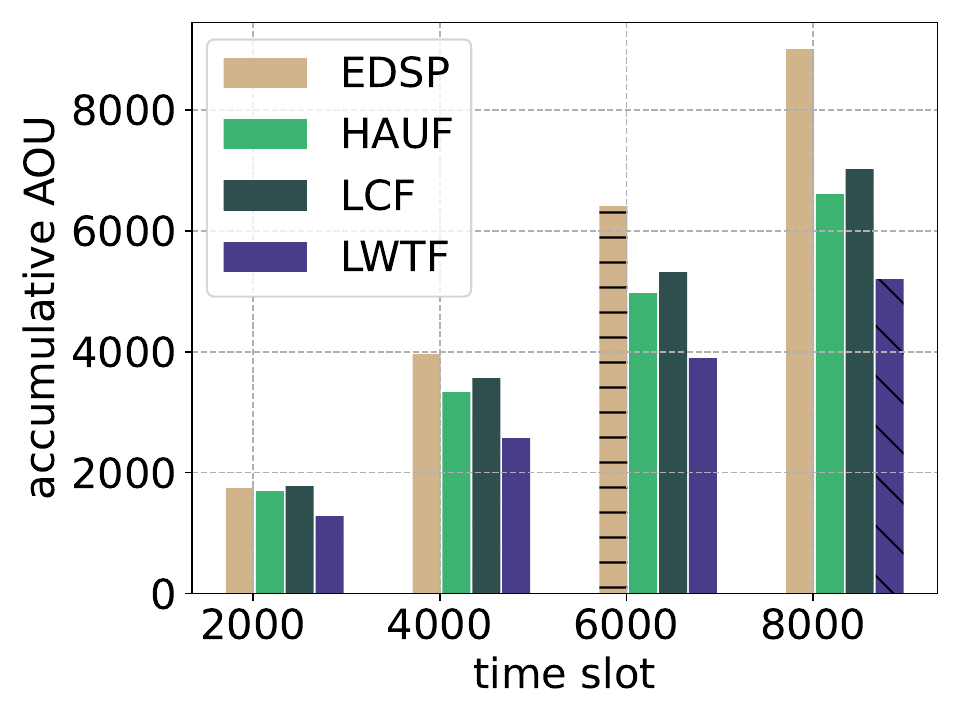}
            \caption{\textsc{Aou} vs. $T$.}
            \label{T}
    \end{minipage}
\end{figure*}

{\color{black}In this section, we demonstrate the performance of \textsc{Esdp} under different scales of scenario settings. 
Fig. \ref{L} and Fig. \ref{R} demonstrate the impact of the scale of the bipartite graph $\mathcal{G}$. Firstly, we observe 
that, whatever the number of the node is, \textsc{Esdp} takes the leading position. Besides, as $|\mathcal{R}|$ becomes larger, 
all the algorithms obtain a relatively larger cumulative \textsc{Aou}. The result is evident because a large cluster can 
provide sufficient computing devices, which leads to jobs being fully served. It is worth noting that, when $|\mathcal{R}|$ 
increases from $60$ to $80$, HAUF achieves a higher \textsc{Aou} than \textsc{Esdp}. The reason of the weak position of \textsc{Esdp} 
is that the solution space $\mathcal{X}$ increases with the node number, and \textsc{Esdp} need a larger time slot length to learn 
the underlying distributions of the computation utilities. Fig. \ref{L} shows that the number of job types, i.e., $|\mathcal{L}|$, 
has a similar impact to $|\mathcal{R}|$ in terms of the performance of \textsc{Esdp}.

Fig. \ref{T} shows that, whatever the parameter settings, \textsc{Esdp} always performs the best, and its performance has a positive 
correlation with the time horizon length $T$. As we have analyzed, a large time horizon provides more chances for \textsc{Esdp} to 
learn the underlying distributions,thereby increasing the reward in the gradient ascent directions.
}
% ======================================================================================================================================================

\section{Related Works}\label{s2}
{\color{black}
Online resource allocation for co-located jobs is always the focus of attention for both industrial and research communities. 
Online algorithms which yield a nice theoretical performance bound can be divided into two categories. 

In the first category, 
the online algorithms are sophistically designed for specific job types, including multi-stage data query and analysis workflows 
(which are organized as DAGs) \cite{8416357,8486340}, service function chains (SFCs) \cite{9154603}, distributed deep neural 
network training jobs \cite{pollux,8486422,9328612,bao2018online,yu2021sum,BSP,narayanan2020heterogeneity}, etc. 
In these works, the algorithms are proposed by formulating combinatorial optimization problems with scenario-oriented constraints, 
and their performance guarantees are provided by the adopted optimization techniques. A typical work is \cite{BSP}, where the 
authors propose an algorithm, named \texttt{SPIN}, with a rounding-based randomized approximation approch, to schedule the 
placement-sensitive Bulk Synchronous Parallel (BSP) jobs. Their design is built on the relaxation of the Gang scheduling 
constraints and the job completion time (JCT) is minimized with linear programming. The authors develop an algorithm which is 
$\mathcal{O} (\ln |\mathcal{M}|)$-approximate with high probability, where $\mathcal{M}$ is the set of computing devices.

In the second category, the job type is not specified, but their theoretical superiority for job co-location and resource contention 
is highlighted. The algorithms are designed with different theoretical basis, including online approximate algorithms 
\cite{8941266,8737612,8917749}, Online Convex Optimization (OCO) techniques \cite{8737465,8737511}, game-theoretical approaches 
\cite{8737370}, Multi-Armed Bandit (MAB) theories and DRL-based algorithms \cite{liang2020data,9488701}, etc. In these works, the 
performance of the proposed algorithms are usually analyzed with approximate ratio, competitive ratio, Price of Anarchy (PoA), 
and regret. A typical recent work is \cite{8737465}. Among these, the most similar work to ours is \cite{8737511}. This work 
presents an online algorithm based on the MAB theories and the OCO techniques, which aims at make online resource allocation decisions 
without knowing future job arrivals according to machine availabilities. The proposed algorithm can achieve 
$\mathcal{O} (\sqrt{T \log \frac{T}{\delta}})$ regret with probability $1 - \delta$ while guaranteeing a small fit for both the 
single-job and multi-job cases over a duration of $T$ time slots. The main differences between this work and ours are summarized 
as follows.
\begin{itemize}
    \item Although \cite{8737511} considers the fluctuated machine service capacities, its system model does not differentiate computing 
    device types. The authors adopt the combinatorial MAB framework to address the resource allocation problem while our algorithm 
    \textsc{Esdp} is built on the AESCB policy. 
    \item In \cite{8737511}, the authors propose an algorithm which has a $\mathcal{O} (\sqrt{T \log \frac{T}{\delta}})$ regret 
    with probability $1 - \delta$ for concave utility functions. By contrast, \textsc{Esdp} has a logarithmic regret $\mathcal{O} (\ln T)$ 
    for linear separatable utilities. Although \textsc{Esdp} has a lower regret in terms of the time slot length $T$, its performance 
    guarantee does suitable for the non-linear cases.
\end{itemize}
}
% ======================================================================================================================================================

\section{Conclusion}\label{s6}
{\color{black}In this paper, we study the multi-server job scheduling problem without knowing the actual processing speed 
distributions apriori. We formulate the problem as a stochastic cumulative overall utility maximization program and cast it into 
the framework of online learning. We propose an online algorithm, termed as \textsc{Esdp}, to learn the underlying processing 
speed distributions and use the exploited statistics to guide the scheduling decisions. \textsc{Esdp} adopts dynamic 
programming to solve several well designed approximated deterministic problems in polynomial time. We prove that \textsc{Esdp} 
has a best possible regret, i.e., $\ln (T)$. The performance of \textsc{Esdp} is also validated with extensive simulations. 
Moreover, extending \textsc{Esdp} to general non-linear utilities might be an interested future research direction. 
}

\ifCLASSOPTIONcompsoc
  \section*{Acknowledgments}
\else
  \section*{Acknowledgment}
\fi
This work was partially supported by the National Science Foundation of China (NSFC) under Grants U20A20173 and 62125206, and 
the Key Research Project of Zhejiang Province under Grant 2022C01145. Schahram Dustdar's work is supported by the Zhejiang 
University Deqing Institute of Advanced technology and Industrilization (ZDATI). 

% ======================================================================================================================================================

\bibliographystyle{IEEEtran}
\bibliography{IEEEabrv,ref.bib}
% ======================================================================================================================================================

\begin{IEEEbiography}
    [{\includegraphics[width=1in,height=1.25in,clip,keepaspectratio]{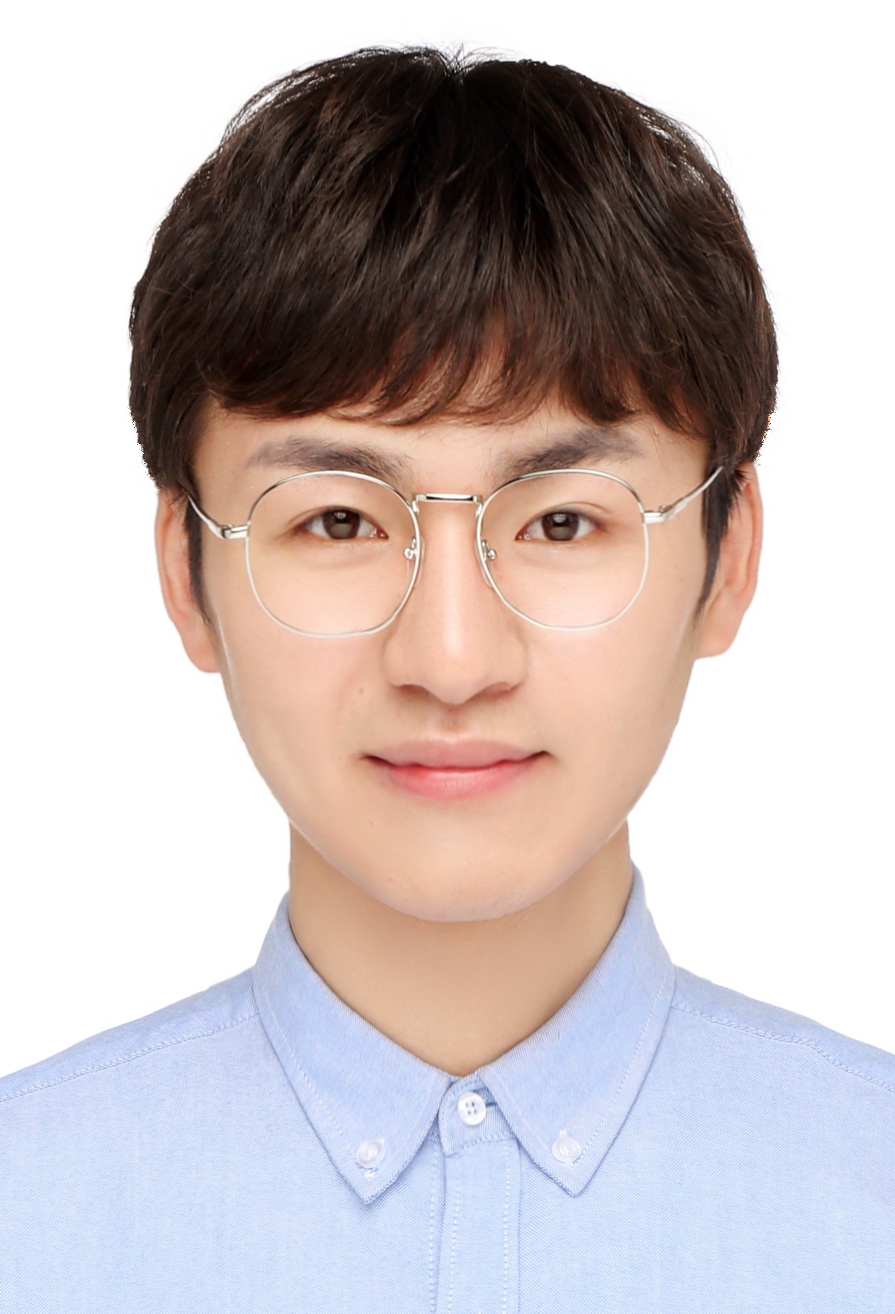}}]{Hailiang Zhao} received the B.S. degree in 
    2019 from the school of computer science and technology, Wuhan University of Technology, Wuhan, China. He is currently pursuing the 
    Ph.D. degree with the College of Computer Science and Technology, Zhejiang University, Hangzhou, China. His research interests include 
    cloud \& edge computing, distributed systems and optimization algorithms. He has published several papers in flagship conferences 
    and journals including IEEE ICWS 2019, IEEE TPDS, IEEE TMC, etc. He has been a recipient of the Best Student Paper Award of IEEE ICWS 2019. 
    He is a reviewer for IEEE TSC and Internet of Things Journal.
\end{IEEEbiography}

\begin{IEEEbiography}
    [{\includegraphics[width=1in,height=1.25in,clip,keepaspectratio]{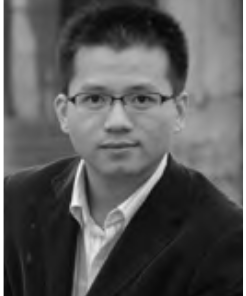}}]{Shuiguang Deng} 
    is currently a full professor at the College of Computer Science and Technology in Zhejiang University, China, 
    where he received a BS and PhD degree both in Computer Science in 2002 and 2007, respectively. He previously 
    worked at the Massachusetts Institute of Technology in 2014 and Stanford University in 2015 as a visiting scholar. 
    His research interests include Edge Computing, Service Computing, Cloud Computing, and Business Process Management. 
    He serves for the journal IEEE Trans. on Services Computing, Knowledge and Information Systems, Computing, and IET 
    Cyber-Physical Systems: Theory \& Applications as an Associate Editor. Up to now, he has published more than 100 
    papers in journals and refereed conferences. In 2018, he was granted the Rising Star Award by IEEE TCSVC. He is 
    a fellow of IET and a senior member of IEEE.
\end{IEEEbiography}

\begin{IEEEbiography}
    [{\includegraphics[width=1in,height=1.25in,clip,keepaspectratio]{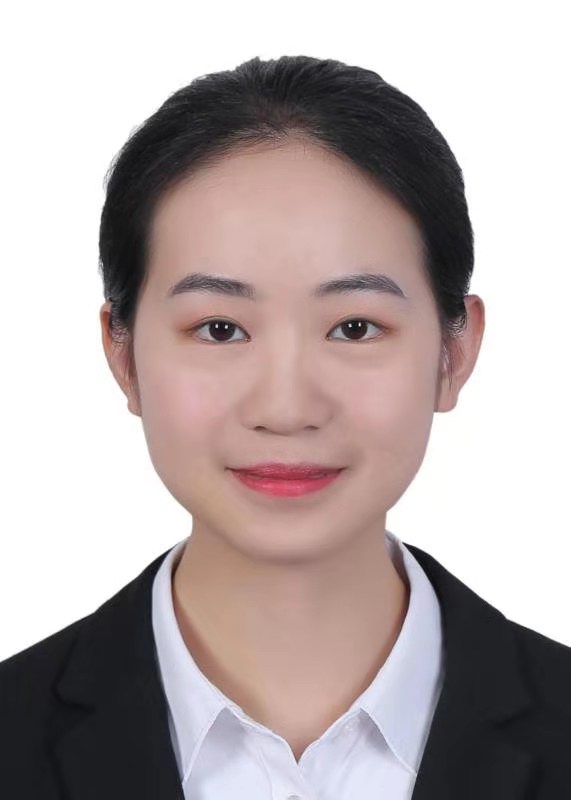}}]{Feiyi Chen} 
    received the B.S. degree in 2021 from the school of computer science and engineering, Sun Yat-sen University (SYSU), Guangzhou, China. 
    She is currently pursuing the master degree with the College of Computer Science and Technology, Zhejiang University, 
    Hangzhou, China. Her research interests include cloud computing, edge computing, and distributed systems.
\end{IEEEbiography}

\begin{IEEEbiography}
    [{\includegraphics[width=1in,height=1.25in,clip,keepaspectratio]{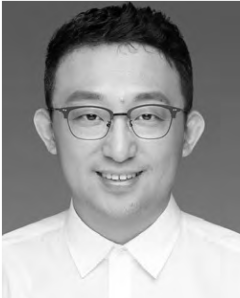}}]{Jianwei Yin} 
    received the Ph.D. degree in computer science from Zhejiang University (ZJU) in 2001. 
    He was a Visiting Scholar with the Georgia Institute of Technology. He is currently a Full Professor 
    with the College of Computer Science, ZJU. Up to now, he has published more than 100 papers in top 
    international journals and conferences. His current research interests include service computing 
    and business process management. He is an Associate Editor of the IEEE Transactions on Services 
    Computing.
\end{IEEEbiography}

\begin{IEEEbiography}[{\includegraphics[width=1in,height=1.25in,clip,keepaspectratio]{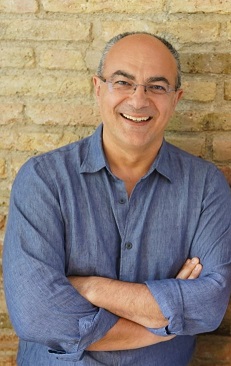}}]{Schahram Dustdar}
    is a Full Professor of Computer Science (Informatics) with a focus on Internet Technologies heading the Distributed 
    Systems Group at the TU Wien. He is founding co-Editor-in-Chief of ACM Transactions on Internet of Things (ACM TIoT) as well as Editor-in-Chief of Computing (Springer). He is an Associate Editor of IEEE Transactions on Services Computing, IEEE Transactions on Cloud Computing, ACM Computing Surveys, ACM Transactions on the Web, and ACM Transactions on Internet Technology, as well as on the editorial board of IEEE Internet Computing and IEEE Computer. Dustdar is recipient of multiple awards: TCI Distinguished Service Award (2021), IEEE TCSVC Outstanding Leadership Award (2018), IEEE TCSC Award for Excellence in Scalable Computing (2019), ACM Distinguished Scientist (2009), ACM Distinguished Speaker (2021), IBM Faculty Award (2012). He is an elected member of the Academia Europaea: The Academy of Europe, where he is chairman of the Informatics Section, as well as an IEEE Fellow (2016), an Asia-Pacific Artificial Intelligence Association (AAIA) President (2021) and Fellow (2021). He is an EAI Fellow (2021) and an I2CICC Fellow (2021). He is a Member of the 2022 IEEE Computer Society Fellow Evaluating Committee (2022).

\end{IEEEbiography}

\begin{IEEEbiography}[{\includegraphics[width=1in,height=1.25in,clip,keepaspectratio]{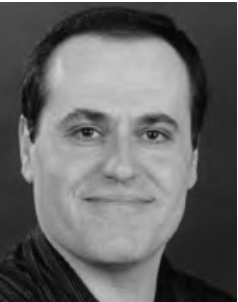}}]{Albert Y. Zomaya}
    is the Peter Nicol Russell Chair Professor of Computer Science and Director of the Centre for Distributed 
    and High-Performance Computing at the University of Sydney. To date, he has published > 600 scientific papers and articles and is (co-)author/editor 
    of > 30 books. A sought-after speaker, he has delivered > 250 keynote addresses, invited seminars, and media briefings. His research interests 
    span several areas in parallel and distributed computing and complex systems. He is currently the Editor in Chief of the ACM Computing Surveys 
    and processed in the past as Editor in Chief of the IEEE Transactions on Computers (2010-2014) and the IEEE Transactions on Sustainable Computing (2016-2020).
    
    Professor Zomaya is a decorated scholar with numerous accolades including Fellowship of the IEEE, the American Association for the Advancement 
    of Science, and the Institution of Engineering and Technology (UK). Also, he is an Elected Fellow of the Royal Society of New South Wales and 
    an Elected Foreign Member of Academia Europaea. He is the recipient of the 1997 Edgeworth David Medal from the Royal Society of New South Wales 
    for outstanding contributions to Australian Science, the IEEE Technical Committee on Parallel Processing Outstanding Service Award (2011), 
    IEEE Technical Committee on Scalable Computing Medal for Excellence in Scalable Computing (2011), IEEE Computer Society Technical Achievement 
    Award (2014), ACM MSWIM Reginald A. Fessenden Award (2017), the New South Wales Premier’s Prize of Excellence in Engineering and Information 
    and Communications Technology (2019), and the Research Innovation Award, IEEE Technical Committee on Cloud Computing (2021). 
  \end{IEEEbiography}

\end{document}